\documentclass[aps,pre,preprint,groupedaddress,floatfix]{revtex4-2}

\usepackage{bm,amsmath,amssymb,amsthm}
\usepackage[dvipdfm]{graphicx}
\usepackage{color}

\newtheorem{theorem}{Theorem}[section]
\newtheorem{lemma}[theorem]{Lemma}

\begin{document}


\title{Enskog and Enskog--Vlasov equations with a slightly modified correlation factor and their H theorem}


\author{Shigeru Takata}
\email[]{takata.shigeru.4a@kyoto-u.ac.jp}
\author{Aoto Takahashi}
\email[]{takahashi.aoto.63c@st.kyoto-u.ac.jp}
\affiliation{Department of Aeronautics and Astronautics, Kyoto University,
Kyoto-daigaku-katsura, Kyoto 615-8540, Japan}


\date{\today}
\begin{abstract}
A novel modification of the original Enskog equation is proposed.
The modification {changes the correlation factor
from a function of density to a functional of density
in a simple form without series structure.}
The proposed modification is {semi-phenomenological 
but is generic} to be adapted to {a given equation of state} for non-ideal gases. 
It is shown that the H-theorem can be established for the Enskog and the Enskog--Vlasov equation with the proposed modification.
\end{abstract}

\keywords{Enskog equation, kinetic theory, dense gas, H theorem.}
\maketitle
\section{Introduction}

Behavior of ideal gases is well described by the Boltzmann equation
for the entire range of the Knudsen number, the ratio of the mean
free path of gas molecules to a characteristic length of the system.
The kinetic theory based on the Boltzmann equation and its model equations have been applied successfully to analyses of various gas flows in low pressure circumstances, micro-scale gas flows, and gas flows caused by the evaporation and/or condensation at the gas-liquid interface. 

The extension of the kinetic theory to non-ideal gases would go back to the dates of Enskog \cite{E72}.
He took account of the displacement effect of molecules in collision integrals for a hard-sphere gas and proposed a kinetic equation that is nowadays called the (original) Enskog equation (OEE).
In the original Enskog equation, there is a correlation factor that
represents an equilibrium correlation function at the contact point of two colliding molecules. 
Although satisfactory outcomes of the OEE, 
such as the dense gas effects on the transport properties,
led to recent developments of numerical algorithms {\cite{MS96,F97,MS97a,WLRZ16}} and their applications to physical problems, e.g., {\cite{F97b,F99,KKW14,WLRZ16,FGLS18,HTT22}},
the intuitive choice of the correlation factor was recognized as causing difficulties in recovering the {Onsager reciprocity in multi-component gases and in establishing the H theorem.
These demerits triggered off} further intensive studies on the foundation of the equation around from late 60's to early 80's, see, e.g.,~\cite{VE73,R78,DVK21} and {references therein.}

Among many efforts in the above-mentioned period, 
Resibois~\cite{R78} succeeded to prove the H theorem, not
for the original but for the modified (or revised) Enskog equation (MEE) \cite{VE73}{.
Resibois's} work motivated further theoretical researches, e.g.,~\cite{BLPT91,MGB18,BB18,BB19,T24}, and the H theorem for the MEE was later extended to include isolated systems \cite{MGB18} and closed systems in contact with a heat bath \cite{T24}; see also the monograph of Dorfman~\emph{et~al.}~\cite{DVK21}.

The MEE has thus offered a satisfactory basis to the theoretical study of dense gases. 
Nevertheless, {the mathematical intricacy has been hindering its further applications} to various fundamental thermal-fluid-dynamic problems{, especially to non-equilibrium boundary-value problems.}
Indeed, to our best knowledge, no numerical results have been reported so far based on the MEE{, except for the case where the density is uniform and accordingly MEE is reduced to OEE
\cite{MS96,MS97}.}
Rather recently, Benilov \& Benilov proposed an alternative {construction of the correlation factor from the semi-phenomenological perspective}, which is more flexible to a given equation of state {(EoS)} for a non-ideal gas \cite{BB18,BB19}. However, their construction {is based} on a series expression taking care of many-body configurations{, like the MEE,} and requires a truncation at a certain order reasonable for both accuracy and computational cost {caused by the rapidly growing complexity}.

In the present work, {we shall focus on a single-component dense gas and 
newly propose a much simpler correlation factor from the semi-phenomenological perspective}
that is free from the series structure and from the many-body configuration.
Moreover, we establish the H theorem for the Enskog equation with the proposed correlation factor. 
Hereinafter, we shall call the Enskog equation with the proposed correlation factor the {Enskog equation with a  slight modification (EESM)} to discriminate from the others, since our proposal is a slight modification of the correlation factor in the OEE.
{Since multi-component systems are not the target of the present work,
we will not discuss the issue related to the Onsager reciprocity.}
The rest of the paper is organized as follows.
First, a generic description of the Enskog equation,
associated notation, and a novel form of the correlation factor
are presented in Sec.~\ref{sec:prbfrm}.
Next it is shown in Sec.~\ref{sec:main} that there is a function
monotonically decreasing in time for the {EESM} and the H-theorem is established for typical physical settings:
(i) a periodic domain, (ii) an isolated domain, and (iii) a closed domain in contact with a heat bath. 
Some technical calculations that are required to obtain the results in Sec.~\ref{sec:main}
are summarized in Appendices~\ref{app:shift}--\ref{sec:KBC}.
Further supplemental discussions on the boundedness of that function
and the extension of the results in Sec.~\ref{sec:main} to the case of the Enskog--Vlasov equation {\cite{G71}} is presented in Appendices~\ref{app:boundedness} and \ref{app:EV}.
The paper is concluded in Sec.~\ref{sec:Conclusion}.

\section{The Enskog equation and the novel correlation factor\label{sec:prbfrm}}

We consider the Enskog equation for a single species dense gas
that is composed of hard sphere molecules with a common diameter $\sigma$ and mass $m$.
Let $D$ be a fixed spatial domain in which the center of gas molecules is confined. 
Let $t$, $\bm{X}$ and $\bm{Y}$, and $\bm{\xi}$
be a time, spatial positions, and a molecular velocity, respectively.
Denoting the one-particle distribution function of gas molecules
by $f(t,\bm{X},\bm{\xi}$), 
the Enskog equation is written as
\begin{subequations}\label{MEE}
\begin{align}
 & \frac{\partial f}{\partial t}+\xi_{i}\frac{\partial f}{\partial X_{i}}=J(f)\equiv J^{G}(f)-J^{L}(f),\quad \mathrm{for\ }\bm{X}\in D,\displaybreak[0]\label{eq:2.1}\\
 & J^{G}(f)\equiv\frac{\sigma^{2}}{m}\int {g(\bm{X}_{\sigma\bm{\alpha}}^{+},\bm{X})f_{*}^{\prime}(\bm{X}_{\sigma\bm{\alpha}}^{+})f^{\prime}(\bm{X})} \notag\\&\qquad\qquad\times V_{\alpha}\theta(V_{\alpha})d\Omega(\bm{\alpha})d\bm{\xi}_{*},\displaybreak[0]\label{eq:2.2}\\
 & J^{L}(f)\equiv\frac{\sigma^{2}}{m}\int {g(\bm{X}_{\sigma\bm{\alpha}}^{-},\bm{X})f_{*}(\bm{X}_{\sigma\bm{\alpha}}^{-})f(\bm{X})} \notag\\&\qquad\qquad\times V_{\alpha}\theta(V_{\alpha})d\Omega(\bm{\alpha})d\bm{\xi}_{*},\label{eq:2.3}
\end{align}
\end{subequations}
\noindent
where $\bm{X}_{\bm{x}}^{\pm}=\bm{X}\pm\bm{x}$,
$\bm{\alpha}$ is a unit vector, 
$d\Omega(\bm{\alpha})$ is a
solid angle element in the direction of $\bm{\alpha}$,
$\theta$ is the Heaviside function
\begin{subequations}
\begin{equation}
\theta(x)=\begin{cases}
1, & x\ge0\\
0, & x<0
\end{cases},
\end{equation}
\noindent
and the following notation convention has been used:
\begin{align}
 & \begin{cases}
{
f(\bm{X})=f(\bm{X},\bm{\xi}),\ f^{\prime}(\bm{X})=f(\bm{X},\bm{\xi}^{\prime})},\\
{
f_{*}(\bm{X}_{\sigma\bm{\alpha}}^{-})=f(\bm{X}_{\sigma\bm{\alpha}}^{-},\bm{\xi}_{*}),\ f_{*}^{\prime}(\bm{X}_{\sigma\bm{\alpha}}^{-})=f(\bm{X}_{\sigma\bm{\alpha}}^{-},\bm{\xi}_{*}^{\prime})},
\end{cases}\label{eq:contf}\displaybreak[0]\\
 & \begin{cases}
{
 \bm{\xi}^{\prime}=\bm{\xi}+V_{\alpha}\bm{\alpha},\quad
 \bm{\xi}_{*}^{\prime}=\bm{\xi}_{*}-V_{\alpha}\bm{\alpha},
 }\\
{ 
 V_{\alpha}=\bm{V}\cdot\bm{\alpha},\quad\bm{V}=\bm{\xi_{*}}-\bm{\xi}.}
 \end{cases}\label{eq:2.5}
\end{align}
\noindent
Here and in what follows, the argument $t$ is often suppressed, unless confusion is anticipated. 
The convention \eqref{eq:contf} will {be applied} only to the quantities that depend on molecular velocity.
{It should be noted that \eqref{MEE} makes sense only 
when the positions $\bm{X}$, $\bm{X}^+_{\sigma\bm{\alpha}}$,
and $\bm{X}^-_{\sigma\bm{\alpha}}$ are all in the domain $D$,
which may restrict the range of integration 
with respect to $\bm{\alpha}$ and $\bm{\xi}_*$.
However, by including the indicator function $\chi_D$
\begin{equation}
\chi_{D}(\bm{X})=\begin{cases}
1, & \bm{X}\in D\\
0, & \mbox{otherwise}
\end{cases},\label{eq:chi_def}
\end{equation}
\noindent
in the definition of $g$ in such a way that
\begin{equation}
g(\bm{X},\bm{Y})=\mathsf{g}(\bm{X},\bm{Y})\chi_D(\bm{X})\chi_D(\bm{Y}),\label{eq:sfg}
\end{equation}}%
\end{subequations}
\noindent
the range of integration in
(\ref{eq:2.2}) and (\ref{eq:2.3}) can be treated as the whole space
of $\bm{\xi}_*$ and all directions of $\bm{\alpha}$, irrespective of the position in the domain $D$.

The factor $\mathsf{g}$ occurring in \eqref{eq:sfg} is generically
assumed to be positive and symmetric with respect to the exchange of two position vectors: $\mathsf{g}(\bm{X},\bm{Y})=\mathsf{g}(\bm{Y},\bm{X})$.
There are some varieties of $\mathsf{g}$ in the literature.
In the case of the OEE, it is assumed to be the pair correlation function
evaluated as a function of the density in the hard-sphere gas in uniform equilibrium.
However, its exact expression is not available \cite{S16} and the Carnahan--Starling approximation \cite{CS69} is often used instead, e.g., \cite{MS96,F97,MS97a,F99,KKW14,MS97,H23}:
\begin{subequations}
\begin{equation}
 \mathsf{g}(\bm{r}_1,\bm{r}_2)
=\frac{1-\phi/2}{(1-\phi)^3},\quad \phi=\frac{\pi\sigma^3}{6m}\rho(\frac{\bm{r}_1+\bm{r}_2}{2}),
\end{equation}
\noindent
where $\rho$ is the density defined by 
\begin{equation}
\rho(t,\bm{X})=\int f(t,\bm{X},\bm{\xi})d\bm{\xi}.
\label{eq:rho_def}
\end{equation}
\noindent
In the case of the MEE, $\mathsf{g}$ is given as 
a series of functionals of the density in the form \cite{VE73,DVK21}
\begin{multline}
\mathsf{g}(\bm{r}_1,\bm{r}_2)
=1+\int \frac{\rho(\bm{r}_3)}{m}\mathcal{V}(\bm{r}_1\bm{r}_2|\bm{r}_3)d\bm{r}_3 \\
  +\frac12\int \frac{\rho(\bm{r}_3)}{m}\frac{\rho(\bm{r}_4)}{m}
   \mathcal{V}(\bm{r}_1\bm{r}_2|\bm{r}_3\bm{r}_4)d\bm{r}_3d\bm{r}_4
+\cdots,\label{eq:g_MEE}
\end{multline}
\noindent
where $\mathcal{V}(\bm{r}_1\bm{r}_2|\bm{r}_3\cdots\bm{r}_l)$ is the sum of all Mayer graphs of $l$ labelled points.
The factor $\mathsf{g}$ in \eqref{eq:g_MEE} is a pair correlation function
that treats the density dependence  
in the hard-sphere gas in nonuniform equilibrium
and reduces to the one in the OEE when the density is uniform.

Although the examples in the previous paragraph are for the hard-sphere gas, 
it is possible to use different forms of $\mathsf{g}$ from a semi-phenomenological perspective.
Indeed, Benilov \& Benilov \cite{BB18,BB19} have recently proposed 
the following alternative form
\begin{multline}
\mathsf{g}(\bm{r},\bm{r}_1)=
1+\sum_{l=1}^\infty c_l\int
\Big[ \prod_{i=2}^{l+1}\prod_{j=i+1}^{l+1}
      \theta(\sigma-|\bm{r}_i-\bm{r}_j|)
\Big]\\
\times
\Big[ \prod_{i=2}^{l+1} \frac{\rho(\bm{r}_i)}{m}
      \theta(\sigma-|\bm{r}-\bm{r}_i|)\theta(\sigma-|\bm{r}_1-\bm{r}_i|)
\Big]\prod_{i=2}^{l+1}d\bm{r}_i,\label{eq:g_BB}
\end{multline}
\end{subequations}
\noindent
where $c_l$'s are constants for adjusting to a given EoS.
The form \eqref{eq:g_BB} involves \eqref{eq:g_MEE} as a special case
and ensures the H theorem as in the case of the MEE.
The semi-phenomenological approach thus has an advantage of flexibility 
in choosing the desired EoS.
However, when using \eqref{eq:g_BB}, the desired EoS can be recovered only approximately, 
since the terms become rapidly complicated and the series practically has to be truncated, leaving the first few terms.

In the present paper, retaining the aforementioned symmetry,
we newly propose the following much simpler form of $\mathsf{g}$:
\begin{subequations}\label{eq:g_prop}\begin{align}
& \mathsf{g}(\bm{X},\bm{Y})=\mathcal{S}(\mathcal{R}(\bm{X}))+\mathcal{S}(\mathcal{R}(\bm{Y})),
\label{eq:g_def}
\\
& \mathcal{R}(\bm{X})=\frac{1}{m}\int_D \rho(\bm{Y})\theta(\sigma-|\bm{Y}-\bm{X}|) d\bm{Y},
\label{eq:R_def}
\end{align}\end{subequations}
\noindent
where $\mathcal{S}$ is a non-negative function
and show that the H theorem is ensured for the Enskog equation
equipped with \eqref{eq:g_prop}.
The specific form of $\mathcal{S}$ will be determined
{in the connection to the EoS of the gas under consideration from the semi-phenomenological perspective.
As is clear from \eqref{eq:R_def}, the key difference from the OEE
is that $\mathsf{g}$ is no longer a function of the density but of the functional $\mathcal{R}$ of density representing the number of molecules in the sphere
with radius $\sigma$ centered at the position under consideration,
which may be interpreted to reflect the exclusion volume effect of a molecule on the probability of the collision partner through $\mathcal{S}$.}

We close this section by listing the definitions
of macroscopic quantities for later convenience.
In addition to the density $\rho$ already given in \eqref{eq:rho_def},
the flow velocity $\bm{v}$ (or $v_i$) and temperature $T$
are defined by
\begin{subequations}
\begin{align}
& v_i=\frac{1}{\rho}\langle \xi_i f \rangle, \label{eq:flow_def}\\
& T=\frac{1}{3R\rho}\langle (\bm{\xi}-\bm{v})^2 f \rangle, \label{eq:temp_def}
\end{align}
\noindent
and the so-called kinetic part of the specific internal energy $e^{(k)}$,
that of the stress tensor $p_{ij}^{(k)}$, and that of the heat-flow vector $\bm{q}^{(k)}$ (or $q_i^{(k)}$), are defined by
\begin{align}
& e^{(k)}=\frac{1}{2\rho}\langle (\bm{\xi}-\bm{v})^2 f \rangle
  (=\frac32RT), \label{eq:kin_inteng_def}\\
& p_{ij}^{(k)}=\langle(\xi_i-v_i)(\xi_j-v_j)f\rangle, \label{eq:kin_stress_def}\\
& q_i^{(k)}=\frac12\langle(\xi_i-v_i)(\bm{\xi}-\bm{v})^2 f\rangle, \label{eq:kin_heatfl_def}
\end{align}
\end{subequations}
\noindent
where $\langle\bullet\rangle=\int \bullet d\bm{\xi}$.

\section{Main results\label{sec:main}}
\subsection{Kinetic part of the H function}

First we shall focus on the so-called kinetic part of the H function
\cite{fn1}
that is defined by
\begin{equation}
\mathcal{H}^{(k)}\equiv\int_{D}\langle f{\ln f} \rangle d\bm{X}.\label{H_kinetic}
\end{equation}
\noindent
The transformation in this small section 
does not require the proposed form
of the correlation factor and can be found in the literature,
e.g., \cite{R78,BLPT91,MGB18,T24}.

Multiply the Enskog equation (\ref{eq:2.1}) by $(1+{\ln f})$
and integrate the result with respect to $\bm{\xi}$ gives
\begin{equation}
\frac{\partial}{\partial t}\langle f{\ln f}\rangle+\frac{\partial}{\partial X_{i}}\langle\xi_{i}f{\ln f}\rangle=\langle J(f){\ln f}\rangle.\label{eq:27}
\end{equation}
\noindent
By further integrating \eqref{eq:27} with respect to $\bm{X}$ over the domain $D$, we have
\begin{equation}
\frac{d}{d t}\mathcal{H}^{(k)}
+\int_D\frac{\partial}{\partial X_{i}}\langle\xi_{i}f{\ln f}\rangle d\bm{X}=\int_{\mathbb{R}^3} \langle J(f){\ln f}\rangle d\bm{X},\label{eq:H^k}
\end{equation}
\noindent
where the range of integration on the right-hand side
has been changed from $D$ to $\mathbb{R}^3$, thanks to the indicator function $\chi_D$ occurring in $g$ [see \eqref{eq:g_def}].
No restriction on the range of spatial integration allows the shift and other operations summarized in Appendix~\ref{app:shift}.

As is explained in Appendix~\ref{app:shift},
the right-hand side of \eqref{eq:H^k} can be transformed into the form that
\begin{align}
 & \int_{\mathbb{R}^3}\langle J(f){\ln f}\rangle d\bm{X}\nonumber \\
= & \frac{\sigma^{2}}{2m}\int\ln\Big(\frac{f_{*}^{\prime}(\bm{X}_{\sigma\bm{\alpha}}^{-})f^{\prime}(\bm{X})}{f_{*}(\bm{X}_{\sigma\bm{\alpha}}^{-})f(\bm{X})}\Big)
g(\bm{X},\bm{X}_{\sigma\bm{\alpha}}^{-}) \notag \\
&\times f(\bm{X})f_{*}(\bm{X}_{\sigma\bm{\alpha}}^{-})V_{\alpha}\theta(V_{\alpha})d\Omega(\bm{\alpha})d\bm{\xi}d\bm{\xi}_{*}d\bm{X}.
\label{eq:Jlnf}
\end{align}
\noindent
Since $x\ln(y/x)\le y-x$ for any $x,y>0$ and the equality holds if and only if $x=y$, we have the estimate that
\begin{equation}
\int_{D}\langle J(f){\ln f}\rangle d\bm{X}\le I(t),
\end{equation}
\noindent
where 
\begin{align}
I(t)&=\frac{\sigma^{2}}{2m}\int g(\bm{X},\bm{X}_{\sigma\bm{\alpha}}^{-})[f_{*}^{\prime}(\bm{X}_{\sigma\bm{\alpha}}^{-})f^{\prime}(\bm{X}) \notag\\&\quad -f(\bm{X})f_{*}(\bm{X}_{\sigma\bm{\alpha}}^{-})] V_{\alpha}\theta(V_{\alpha})d\Omega(\bm{\alpha})d\bm{\xi}d\bm{\xi}_{*}d\bm{X},\label{eq:I_def}
\end{align}
\noindent
and the equality holds if and only if $I(t)=0$ or equivalently
\begin{equation}
f_{*}^{\prime}(\bm{X}_{\sigma\bm{\alpha}}^{-})f^{\prime}(\bm{X})-f(\bm{X})f_{*}(\bm{X}_{\sigma\bm{\alpha}}^{-})=0.
\label{eq:equilibrium}
\end{equation}
\noindent
It should be remarked that, 
as is explained in Appendix~\ref{app:ColEnt},
$I(t)$ is eventually reduced to
\begin{align}
I(t)=-\frac{\sigma^{2}}{m}&\int g(\bm{X},\bm{X}_{\sigma\bm{\alpha}}^+)\rho(\bm{X})\rho(\bm{X}_{\sigma\bm{\alpha}}^{+}) \notag\\&\quad\times 
\bm{v}(\bm{X})\cdot\bm{\alpha}d\Omega(\bm{\alpha})d\bm{X}.
\label{eq:I_final}
\end{align}

\subsection{Collisional part of the H function}

Next introduce the following function
\begin{equation}
\mathcal{H}^{(c)}(t)=\int_D \rho(\bm{X}) [\int_0^{\mathcal{R}(\bm{X})}
\mathcal{S}(x)dx ] d\bm{X}.\label{eq:H^c}
\end{equation}
\noindent
We shall show below that its time derivative is related to $I(t)$.
Namely, using a concise notation $r=|\bm{Y}-\bm{X}|$,
\begin{widetext}
\begin{align}
\frac{d}{dt}\mathcal{H}^{(c)}
=&\frac{d}{dt}\int_D \rho(\bm{X})[\int_0^{\mathcal{R}(\bm{X})}
\mathcal{S}(x)dx ] d\bm{X} \notag \displaybreak[0]\\
=&\int_D\{ \frac{\partial\rho(\bm{X})}{\partial t}
 \int_0^{\mathcal{R}(\bm{X})}\mathcal{S}(x)dx 
+\rho(\bm{X})\frac{\partial\mathcal{R}(\bm{X})}{\partial t}
\mathcal{S}(\mathcal{R}(\bm{X}))\}d\bm{X} \notag \displaybreak[0]\\
=&\int_D
  \{ \frac{\partial\rho(\bm{X})}{\partial t}
  \int_0^{\mathcal{R}(\bm{X})}\mathcal{S}(x)dx
+ \frac{\rho(\bm{X})}{m}[\int_D\frac{\partial\rho(\bm{Y})}{\partial t}
  \theta(\sigma-r)d\bm{Y}]
  \mathcal{S}(\mathcal{R}(\bm{X}))\}d\bm{X} \notag \displaybreak[0]\\
=&\int_D \frac{\partial\rho(\bm{X})}{\partial t}
  [\int_0^{\mathcal{R}(\bm{X})}\mathcal{S}(x)dx
  +\int_D\frac{\rho(\bm{Y})}{m}\theta(\sigma-r)\mathcal{S}(\mathcal{R}(\bm{Y}))d\bm{Y}]d\bm{X} \notag \displaybreak[0]\\
=&-\int_D \frac{\partial\rho{v}_i(\bm{X})}{\partial X_i}
  [\int_0^{\mathcal{R}(\bm{X})}\mathcal{S}(x)dx
  +\int_D\frac{\rho(\bm{Y})}{m}\theta(\sigma-r)\mathcal{S}(\mathcal{R}(\bm{Y}))d\bm{Y}]d\bm{X} \notag \displaybreak[0]\\
=&-\int_D \frac{\partial}{\partial X_i}
  \{\rho{v}_i(\bm{X})[\int_0^{\mathcal{R}(\bm{X})}\mathcal{S}(x)dx
  +\int_D\frac{\rho(\bm{Y})}{m}\theta(\sigma-r)\mathcal{S}(\mathcal{R}(\bm{Y}))d\bm{Y}]\}d\bm{X}\notag \\
& +\int_D \rho{v}_i(\bm{X})
  [\frac{\partial\mathcal{R}(\bm{X})}{\partial X_i}
  \mathcal{S}(\mathcal{R}(\bm{X}))
  +\int_D\frac{\rho(\bm{Y})}{m}\frac{\partial}{\partial X_i}\theta(\sigma-r)\mathcal{S}(\mathcal{R}(\bm{Y}))d\bm{Y}]d\bm{X} \notag \displaybreak[0]\\
=&-\int_D \frac{\partial}{\partial X_i}
  \{\rho{v}_i(\bm{X})[\int_0^{\mathcal{R}(\bm{X})}\mathcal{S}(x)dx
  +\int_D\frac{\rho(\bm{Y})}{m}\theta(\sigma-r)\mathcal{S}(\mathcal{R}(\bm{Y}))d\bm{Y}]\}d\bm{X}\notag \\
& +\int_D \rho{v}_i(\bm{X})
  \{\int_D\frac{\rho(\bm{Y})}{m}\frac{Y_i-X_i}{|\bm{Y}-\bm{X}|}\delta(\sigma-|\bm{Y}-\bm{X}|)[\mathcal{S}(\mathcal{R}(\bm{X}))+\mathcal{S}(\mathcal{R}(\bm{Y}))]d\bm{Y}\}d\bm{X} \notag \displaybreak[0]\\
=&-\int_D \frac{\partial}{\partial X_i}
  \{\rho{v}_i(\bm{X})[\int_0^{\mathcal{R}(\bm{X})}\mathcal{S}(x)dx
  +\int_D\frac{\rho(\bm{Y})}{m}\theta(\sigma-r)\mathcal{S}(\mathcal{R}(\bm{Y}))d\bm{Y}]\}d\bm{X}\notag \\
& +\frac{\sigma^2}{m} \int \rho{v}_i(\bm{X})
  \{\int\rho(\bm{X}_{\sigma\bm{\alpha}}^+)\alpha_i[\mathcal{S}(\mathcal{R}(\bm{X}))+\mathcal{S}(\mathcal{R}(\bm{X}_{\sigma\bm{\alpha}}^+))]\chi_D(\bm{X})\chi_D(\bm{X}_{\sigma\bm{\alpha}}^+)d\Omega(\bm{\alpha})\}d\bm{X} \notag \displaybreak[0]\\
=&-\int_D \frac{\partial}{\partial X_i}
  \{\rho{v}_i(\bm{X})[\int_0^{\mathcal{R}(\bm{X})}\mathcal{S}(x)dx
  +\int_D\frac{\rho(\bm{Y})}{m}\theta(\sigma-r)\mathcal{S}(\mathcal{R}(\bm{Y}))d\bm{Y}]\}d\bm{X}\notag \\
& +\frac{\sigma^2}{m} \int \rho(\bm{X}){v}_i(\bm{X})
  \rho(\bm{X}_{\sigma\bm{\alpha}}^+)\alpha_i
  g(\bm{X},\bm{X}_{\sigma\bm{\alpha}}^+)d\Omega(\bm{\alpha}) d\bm{X} \notag \displaybreak[0]\\
=&-\int_D \frac{\partial}{\partial X_i}
  \{\rho{v}_i(\bm{X})[\int_0^{\mathcal{R}(\bm{X})}\mathcal{S}(x)dx
  +\int_D\frac{\rho(\bm{Y})}{m}\theta(\sigma-r)\mathcal{S}(\mathcal{R}(\bm{Y}))d\bm{Y}]\}d\bm{X} -I(t).
\end{align}
\end{widetext}
\noindent
As the result, the time derivative of the sum
$\mathcal{H}\equiv\mathcal{H}^{(k)}+\mathcal{H}^{(c)}$
is found to satisfy the inequality that
\begin{subequations}
\begin{equation}
\frac{d\mathcal{H}}{dt}
=\frac{d}{dt}(\mathcal{H}^{(k)}+\mathcal{H}^{(c)})
\le 
\int_{\partial D} (H_i^{(k)}+H_i^{(c)})n_i dS,
\label{eq:H_inequality}
\end{equation}
\noindent
where
\noindent
$\bm{n}$ (or $n_i$) is the inward unit normal to the boundary,
$dS$ is the surface element of the boundary $\partial D$, and
\begin{align}
H_i^{(k)}=&\langle \xi_i f\ln f\rangle,\\
H_i^{(c)}=&\rho(\bm{X}){v}_i(\bm{X})[\int_0^{\mathcal{R}(\bm{X})}\mathcal{S}(x)dx \notag \\
& +\int_D\frac{\rho(\bm{Y})}{m}\theta(\sigma-|\bm{Y}-\bm{X}|)\mathcal{S}(\mathcal{R}(\bm{Y}))d\bm{Y}].
\end{align}
\end{subequations}
\noindent
The inequality \eqref{eq:H_inequality} 
is the H theorem for the {EESM}.

\subsection{Monotonicity in three typical cases\label{sec:mono}}
We shall discuss the more details about \eqref{eq:H_inequality}
for three typical cases: the domain $D$ is three dimensional and 
is (i) periodic {\cite{fn:periodic}}, (ii) surrounded by the specular reflection boundary,
and (iii) surrounded by the impermeable surface of a heat bath
with a uniform constant temperature $T_w$. 

First consider the case (i).
In this case, the surface integrals over $\partial D$ vanish,
thanks to the periodicity, and accordingly we have
\begin{equation}
\frac{d\mathcal{H}}{dt}\le 0,\label{eq:H_ineq_period}
\end{equation}
and the equality holds if and only if \eqref{eq:equilibrium} is satisfied.

Next consider the case (ii). 
Then $H_i^{(k)} n_i=0$ on $\partial D$, since $\xi_in_i f\ln f$ is odd with respect to $\xi_i n_i$. 
Noting that the specular reflection boundary is impermeable, 
$H_i^{(c)}n_i=0$ holds as well because of $v_in_i=0$ on $\partial D$ by
the impermeability.
As the result, we have again
\begin{equation}
\frac{d\mathcal{H}}{dt}\le 0,\label{eq:H_ineq_spec}
\end{equation}
and the equality holds if and only if \eqref{eq:equilibrium} is satisfied.

Finally, consider the case (iii).
Because of the impermeability, $v_in_i=0$ and thus
$H_i^{(c)}n_i=0$ on the boundary.
Then, thanks to the Darrozes--Guiraud inequality (Lemma~\ref{lem:(Darrozes--Guiraud)}) \cite{DG66,C88,S07},
it holds that
\begin{subequations}\begin{equation}
\int_{\partial D}\langle\xi_in_i(\bm{X}) f\ln\frac{f}{f_{w}}\rangle dS(\bm{X})\le0,
\end{equation}
\noindent
where 
\begin{equation}
f_{w}=\frac{C}{(2\pi RT_{w})^{3/2}}\exp(-\frac{\bm{\xi}^{2}}{2RT_{w}}),\label{eq:fw}
\end{equation}\end{subequations}
\noindent
with $C(>0)$ is an arbitrary constant. 
Since $T_w(>0)$ is a constant,
the multiplication of $\ln f_w$ with \eqref{eq:2.1}
and its integration with respect to $\bm{X}$ over the domain $D$ leads to
\begin{align}
&\frac{d}{dt}\int_D \langle f\ln f_w \rangle d\bm{X}
-\int_{\partial D}\langle \xi_i n_i f\ln f_w \rangle dS \notag\\
&=\int_D \langle  J(f) \ln f_w \rangle d\bm{X},
\end{align}
\noindent
where the right-hand side vanishes
as is explained in Appendix~\ref{app:ConservationLaws}
[see \eqref{eq:totalmom_eng_transp}]. 
Hence, \eqref{eq:H_inequality} is recast as
\begin{subequations}
\begin{equation}
\frac{d\mathcal{F}}{dt}\le 0,
\end{equation}
\noindent
where
\begin{align}
\mathcal{F}
=&RT_w(\int_D \langle f \ln\frac{f}{f_w}\rangle d\bm{X}+\mathcal{H}^{(c)})
\notag\\
=&RT_w\mathcal{H}+\int_D \langle \frac12\bm{\xi}^2 f\rangle d\bm{X}
+\mathrm{const}.,\label{eq:freeEng}
\end{align}\end{subequations}
\noindent
and the equality holds again if and only if \eqref{eq:equilibrium} is satisfied. 
\cite{fn:freeEng}

{Incidentally, it is readily seen that $\mathcal{H}^{(c)}$
is the counterpart of $\int_\Omega Q(n) d\vec{r}$ in \cite{GG80} multiplied with $m$, where $\Omega$ is a periodic domain, and that $\mathcal{H}^{(c)}$ 
indeed satisfies the \emph{compatibility condition with the thermodynamics}, (3.6) in \cite{GG80} [$\epsilon_{ij}$ in (3.6) should read $\epsilon_{ij}^{-1}$].
Therefore, the present result for the periodic domain is consistent with Theorem III.4 in \cite{GG80}.
Moreover, the extension of Theorem III.4 in \cite{GG80} to the other two cases
considered in the present subsection can be made by simply including the contribution
from $H_i^{(k)}$ and $H_i^{(c)}$ in the discussion there.
It is seen that the same \emph{compatibility condition} applies to these two cases as well
in order to ensure the H theorem. The details are omitted here.}

\subsection{The Maxwellian at the stationary state for the three typical cases\label{sec:Maxwellian}}

The condition \eqref{eq:equilibrium} is rewritten as
\begin{equation}
 \ln f         (\bm{X})
+\ln f_*       (\bm{X}^-_{\sigma\bm{\alpha}})
=\ln f^\prime  (\bm{X})
+\ln f_*^\prime(\bm{X}^-_{\sigma\bm{\alpha}}),
\end{equation}
\noindent
which implies that such a $\ln f$ is the summational invariant
and is restricted to the form
\begin{subequations}\label{eq:Max}
\begin{align}
 f(\bm{X})
=&\frac{\rho(\bm{X})}{(2\pi RT)^{3/2}}
\exp(-\frac{[\bm{\xi}-\bm{v}(\bm{X})]^2}{2RT}), \\
\bm{v}(\bm{X})=&\bm{u}+\bm{X}\times\bm{\omega},
\end{align}
\end{subequations}
\noindent
see, e.g., \cite{TT24,MGB18}.
Here, it should be noted that $T$, $\bm{u}$, and $\bm{\omega}$ are independent of $\bm{X}$ and that
$\bm{u}$ and $\bm{\omega}$ represent
the translational and the angular velocity of the flow, respectively.

In the case (i), the periodic condition prohibits 
the rotational mode of velocity, 
so that \eqref{eq:Max} is reduced to
\begin{equation}
 f(\bm{X})
=\frac{\rho(\bm{X})}{(2\pi RT)^{3/2}}
\exp(-\frac{(\bm{\xi}-\bm{u})^2}{2RT}),
\label{eq:Max_period}
\end{equation}
\noindent
where $T(>0)$ and $\bm{u}$ are uniform scalar and vector
that are determined in accordance 
with the total momentum and energy, thanks to the conservation laws
(see Appendix~\ref{app:ConservationLaws}). 

In the case (ii), the specular reflection does not allow
the translational velocity $\bm{u}$, so that
\eqref{eq:Max} is reduced to
\begin{equation}
 f(\bm{X})
=\frac{\rho(\bm{X})}{(2\pi RT)^{3/2}}
\exp(-\frac{(\bm{\xi}-\bm{X}\times \bm{\omega})^2}{2RT}),\label{eq:RigidRotation}
\end{equation}
\noindent
where a non-zero $\bm{\omega}$ is admitted only when
the shape of the boundary is symmetric around $\bm{\omega}$.
Equation~\eqref{eq:RigidRotation} represents a rigid rotation of the fluid.
Moreover, $T>0$ and $\bm{\omega}$ are again a uniform scalar and vector
that are determined in accordance 
with the total momentum and energy, thanks to the conservation law.
If the shape of the boundary is not symmetric, the total momentum is not conserved and finally the angular velocity vanishes irrespective of the initial state due to the impermeability of the boundary.

Finally, in the case (iii), \eqref{eq:Max} is reduced to
\begin{equation}
 f(\bm{X})
=\frac{\rho(\bm{X})}{(2\pi RT_w)^{3/2}}
\exp(-\frac{\bm{\xi}^2}{2RT_w}),
\end{equation}
\noindent
since the Darrouse--Guiraud inequality (Lemma~\ref{lem:(Darrozes--Guiraud)}) forces the gas to be at rest and at the same temperature as the heat bath on the boundary.

\subsection{Relation to the equation of state}

As is well known, 
the stress tensor $p_{ij}$ in the case of the Enskog equation
is expressed
by the sum of the kinetic and the collisional contribution,
$p_{ij}^{(k)}$ and $p_{ij}^{(c)}$;
see \eqref{eq:kin+col}, \eqref{eq:kin_stress_def}, \eqref{eq:col_stress}, and Appendix~\ref{app:ConservationLaws}. 
Since the pressure $p$ is defined as the one third of the trace of the stress tensor, it is expressed as
\begin{align}
p=&\frac13 (p_{ii}^{(k)}+p_{ii}^{(c)}) \notag \\
 =&\rho RT
 +\frac{\sigma^{2}}{6m}\int {\int_{0}^{\sigma}}
   V_{\alpha}^{2}\theta(V_{\alpha}) g(\bm{X}_{\lambda\bm{\alpha}}^{+},\bm{X}_{(\lambda-\sigma)\bm{\alpha}}^{+}) \notag \\
  &\quad \times f_{*}(\bm{X}_{(\lambda-\sigma)\bm{\alpha}}^{+})f(\bm{X}_{\lambda\bm{\alpha}}^{+}){d\lambda} d\Omega(\bm{\alpha})d\bm{\xi}_{*}d\bm{\xi}.
   \label{eq:pressure}
\end{align}

Now consider the infinite expanse of the dense gas in the uniform equilibrium state. In the case, the last integral is reduced to
{$(\sigma^2/6)(\rho^2 RT/m) \int_0^\sigma \int g(\bm{X}_{\lambda\bm{\alpha}}^{+},\bm{X}_{(\lambda-\sigma)\bm{\alpha}}^{+}) d\Omega(\bm{\alpha}) d\lambda$.}
In the case of the proposed correlation factor \eqref{eq:g_def} with \eqref{eq:R_def},
{$g$ becomes constant such that
\begin{equation}
 g(\bm{X}_{\lambda\bm{\alpha}}^{+},\bm{X}_{(\lambda-\sigma)\bm{\alpha}}^{+})
=2\mathcal{S}(2b\rho),
\quad
b=\frac{2\pi}{3}\frac{\sigma^3}{m},\label{eq:gR_simplified}
\end{equation}
\noindent
since the density is constant.
As the result, \eqref{eq:pressure} is reduced to the following EoS for the gas under consideration:}
\begin{equation}
p=\rho RT (1+2b\rho \mathcal{S}(2b\rho)).
\end{equation}

Now we shall present the specific form of $\mathcal{S}$
for two typical equations of state 
in the literature:
the van der Waals equation of state \cite{vdW} for non-attractive molecules
\begin{equation}
p=\frac{\rho RT}{1-b\rho}=\rho RT(1+\frac{b\rho}{1-b\rho}),
\label{eq:EoSvdW}
\end{equation}
and the Carnahan--Starling equation of state \cite{CS69}
\begin{equation}
p=\rho RT\frac{1+\eta+\eta^2-\eta^3}{(1-\eta)^3}
 =\rho RT(1+\frac{4\eta-2\eta^2}{(1-\eta)^3}),
\label{eq:EoSCS}
\end{equation}
\noindent
where $\eta={b\rho}/{4}$.
It is readily seen that the appropriate form of $\mathcal{S}$  
for the van der Waals equation of state for non-attractive molecules
is
\begin{equation}
\mathcal{S}(x)=\frac{1}{2-x}, \label{eq:SvdW}
\end{equation}
\noindent
while that for the Carnahan--Starling equation of state is
\begin{equation}
 \mathcal{S}(x)=\frac{16(16-x)}{(8-x)^3}. \label{eq:SCS}
\end{equation}

It should be noted that
the collision term of the Enskog equation is not responsible 
for the attractive part of the {EoS}.
The attractive part is to be recovered by the Vlasov term of the Enksog--Vlasov equation; see Appendix~\ref{app:EV} and, e.g., \cite{G71,FGLS18}.
Therefore the form \eqref{eq:SvdW} applies to the {EESM}
with the Vlasov term for the \emph{full} version of the van der Waals fluids as well (see Appendix~\ref{app:EV}).

\section{Conclusion\label{sec:Conclusion}}

In the present paper, a novel form of the correlation factor in the collision integral of the Enskog equation has been proposed. The new factor is a slight modification {of the counterpart of the original Enskog equation from a function to a functional of the density and still keeps a simplicity without a series structure.} 
The H theorem has been established for the Enskog equation with the proposed modification.
The function monotonically decreasing in time
has been presented for three typical cases:
the domain is (i) periodic, (ii) surrounded by the specular reflection boundary, and (iii) surrounded by the impermeable surface of a resting heat bath. 
For these cases, that function {is shown to be bounded} in Appendix~\ref{app:boundedness}
for the van der Waals (without the attractive part) and the Carnahan--Starling equation of state, if it is initially bounded.
These results, i.e., the H theorem, the monotonically decreasing function, and its boundedness, have also been extended to the Enskog--Vlasov equation
in Appendix~\ref{app:EV}.

The specific form of the novel correlation factor 
is shown to be determined in accordance with the {EoS}
(more precisely the repulsive part of the {EoS})
under consideration. 
As two typical examples,
the form in the cases of the van der Waals
and the Carnahan--Starling equation of state
have been presented as well.

{Although simple kinetic models at the level of relaxation model have been proposed and used for the dense gas in the literature, e.g., \cite{DSB96,MTHT23,SSGZ23,MT24}, the present result is expected to activate further numerical applications of the Enskog equation.
Indeed,} the novel correlation factor is so simple to be implemented in the numerical schemes for the OEE in the literature {\cite{MS96,F97,MS97a,WLRZ16}}.
Hence, by a slight modification, those schemes can be used for the {EESM} and put the numerics {of the Enskog equation} on the firm ground of the {thermodynamical} consistency. 

{Finally, although not discussed in the present paper, the fluid-dynamic limit of the EESM recovers the same transport properties as that of the OEE at the level of the Navier--Stokes--Fourier equations, provided that the corresponding EoS is common. The details will be reported elsewhere. Numerical application of the EESM to thermo-fluid phenomena is a natural direction of research stemming from the present work. Another natural direction is to examine the feasibility of the EESM in the case of multi-component systems, in which the Onsager reciprocity should be properly recovered as often mentioned in the literature, e.g., \cite{VE73,DVK21}. This topic, while important, should be the subject of future work.}

\appendix
\section{Shift and other operations \label{app:shift}}
We summarize the shift and other standard operations
that are used in the transformations of the collision integral.

There are three types of operation that
are standard in the case of the Boltzmann equation as well:
\begin{description}
\item [{(I)}] to exchange the letters $\bm{\xi}$ and $\bm{\xi}_{*}$; 
\item [{(II)}] to reverse the direction of $\bm{\alpha}$;
\item [{(III)}] to change the integration variables from $(\bm{\xi},\bm{\xi}_{*},\bm{\alpha})$
to $(\bm{\xi}^{\prime},\bm{\xi}_{*}^{\prime},\bm{\alpha})$ and then to
change the letters $(\bm{\xi}^{\prime},\bm{\xi}_{*}^{\prime})$ to
$(\bm{\xi},\bm{\xi}_{*})$.
\end{description}

First, by (III) and (II),
\begin{equation}
\int{\varphi}(\bm{X}) J^{G}(f)d\bm{\xi}=\int{\varphi}^{\prime}(\bm{X})J^{L}(f)d\bm{\xi},\label{eq:4.1}
\end{equation}
\noindent
holds for any ${\varphi}(\bm{X},\bm{\xi})$.
Hence, we have
\begin{align}
 & \langle\varphi(\bm{X})J(f)\rangle\nonumber \\
=& \frac{\sigma^{2}}{m}
   \int[{\varphi}^\prime(\bm{X})-{\varphi}(\bm{X})]g(\bm{X}_{\sigma\bm{\alpha}}^{-},\bm{X})f_{*}(\bm{X}_{\sigma\bm{\alpha}}^{-})f(\bm{X}) \notag\\ &\qquad\times V_{\alpha}\theta(V_{\alpha})d\Omega(\bm{\alpha})d\bm{\xi}_{*}d\bm{\xi}.\label{eq:mom_J}
\end{align}

Next by integrating \eqref{eq:mom_J} with respect to $\bm{X}$ over the domain $D$, we have
\begin{align}
 & \int_D\langle \varphi(\bm{X})J(f) \rangle d\bm{X}\nonumber \\
=& \frac{\sigma^{2}}{m}
   \int_{\mathbb{R}^3}\int[{\varphi}^\prime(\bm{X})-{\varphi}(\bm{X})]g(\bm{X}_{\sigma\bm{\alpha}}^{-},\bm{X})f_{*}(\bm{X}_{\sigma\bm{\alpha}}^{-})f(\bm{X}) \notag\\ &\qquad\times V_{\alpha}\theta(V_{\alpha})d\Omega(\bm{\alpha})d\bm{\xi}_{*}d\bm{\xi}d\bm{X}.\label{eq:mom_J1}
\end{align}
\noindent
Here the domain of integration with respect to $\bm{X}$
has been changed from $D$ to $\mathbb{R}^3$,
thanks to the factor $\chi_D$ occurring in $g$; see \eqref{eq:g_def}.
This change also allows $\bm{\alpha}$ to take all the directions.
Now consider the shift operation by $\sigma\bm{\alpha}$
followed by the operations (II) and (I).
Then, the right-hand side of \eqref{eq:mom_J1} is recast as
\begin{align}
\frac{\sigma^{2}}{m}
   &\int_{\mathbb{R}^3}\int[{\varphi}^\prime_{*}(\bm{X}_{\sigma\bm{\alpha}}^-)-{\varphi}_{*}(\bm{X}_{\sigma\bm{\alpha}}^-)]g(\bm{X},\bm{X}_{\sigma\bm{\alpha}}^{-})\notag\\ &\qquad\times f(\bm{X})f_{*}(\bm{X}_{\sigma\bm{\alpha}}^-) V_{\alpha}\theta(V_{\alpha})d\Omega(\bm{\alpha})d\bm{\xi}d\bm{\xi}_{*}d\bm{X}.\label{eq:mom_J2}
\end{align}
\noindent
and thus, we have
\begin{align}
 & \int_D\langle \varphi(\bm{X})J(f) \rangle d\bm{X}\notag \\
=& \frac{\sigma^{2}}{2m}
   \int_{\mathbb{R}^3}\int[{\varphi}^\prime(\bm{X})+{\varphi}^\prime_{*}(\bm{X}_{\sigma\bm{\alpha}}^-)-{\varphi}_{*}(\bm{X}_{\sigma\bm{\alpha}}^-)-{\varphi}(\bm{X})] \notag\\
   & \times g(\bm{X}_{\sigma\bm{\alpha}}^{-},\bm{X})f_{*}(\bm{X}_{\sigma\bm{\alpha}}^{-})f(\bm{X})V_{\alpha}\theta(V_{\alpha})d\Omega(\bm{\alpha})d\bm{\xi}_{*}d\bm{\xi}d\bm{X}.\label{eq:mom_Jsum}
\end{align}

The substitution of $\varphi=\ln f$ in \eqref{eq:mom_Jsum} yields
\begin{align}
 & \int_D \langle J(f) \ln f \rangle d\bm{X}\notag \\
=& \frac{\sigma^{2}}{2m}
   \int_{\mathbb{R}^3}\int \ln\frac{f^\prime_{*}(\bm{X}_{\sigma\bm{\alpha}}^-) f^\prime(\bm{X})}{f_{*}(\bm{X}_{\sigma\bm{\alpha}}^-) f(\bm{X})} g(\bm{X}_{\sigma\bm{\alpha}}^{-},\bm{X}) \notag\\
   &\quad \times f_{*}(\bm{X}_{\sigma\bm{\alpha}}^{-})f(\bm{X})V_{\alpha}\theta(V_{\alpha})d\Omega(\bm{\alpha})d\bm{\xi}_{*}d\bm{\xi}d\bm{X},\label{eq:mom_Jlog}
\end{align}
\noindent
which is the form of \eqref{eq:Jlnf}.

\section{Collisional contributions to entropy generation and transport properties\label{app:ColEntTransp}}

\subsection{Collisional contributions to the entropy generation\label{app:ColEnt}}

The transformation of $I(t)$ from \eqref{eq:I_def} to \eqref{eq:I_final}
has been done by the repeated operations of the shift, (II), and (III):
\begin{align}
&I(t) \notag\\
= & -\frac{\sigma^{2}}{2m}\int g(\bm{X},\bm{X}_{\sigma\bm{\alpha}}^{-})f_{*}^{\prime}(\bm{X}_{\sigma\bm{\alpha}}^{-})f^{\prime}(\bm{X})  \notag\\ 
 &\qquad\qquad\times V_{\alpha}^{\prime}\theta(-V_{\alpha}^{\prime}) d\Omega(\bm{\alpha})d\bm{\xi}d\bm{\xi}_{*}d\bm{X}\nonumber  \\
 & -\frac{\sigma^{2}}{2m}\int g(\bm{X},\bm{X}_{\sigma\bm{\alpha}}^{-})f(\bm{X})f_{*}(\bm{X}_{\sigma\bm{\alpha}}^{-}) \notag\\
 &\qquad\qquad\times V_{\alpha}\theta(V_{\alpha}) d\Omega(\bm{\alpha})d\bm{\xi}d\bm{\xi}_{*}d\bm{X}\displaybreak[0]\nonumber \\
= & \frac{\sigma^{2}}{2m}\int g(\bm{X},\bm{X}_{\sigma\bm{\alpha}}^{-})f(\bm{X})f_{*}(\bm{X}_{\sigma\bm{\alpha}}^{-}) \notag\\
&\qquad\qquad\times [(\bm{\xi}-\bm{\xi}_{*})\cdot\bm{\alpha}] d\Omega(\bm{\alpha})d\bm{\xi}d\bm{\xi}_{*}d\bm{X}\displaybreak[0]\nonumber \\
= & \frac{\sigma^{2}}{2m}\int g(\bm{X},\bm{X}_{\sigma\bm{\alpha}}^{-})\rho(\bm{X})\rho(\bm{X}_{\sigma\bm{\alpha}}^{-}) \notag\\&\qquad\quad\times \bm{v}(\bm{X})\cdot\bm{\alpha}d\Omega(\bm{\alpha})d\bm{X}\nonumber \\
 &-\frac{\sigma^{2}}{2m}\int g(\bm{X},\bm{X}_{\sigma\bm{\alpha}}^{-})\rho(\bm{X})\rho(\bm{X}_{\sigma\bm{\alpha}}^{-}) \notag\\&\qquad\quad\times \bm{v}(\bm{X}_{\sigma\bm{\alpha}}^{-})\cdot\bm{\alpha}d\Omega(\bm{\alpha})d\bm{X}\displaybreak[0]\nonumber \\
= & \frac{\sigma^{2}}{2m}\int g(\bm{X},\bm{X}_{\sigma\bm{\alpha}}^{-})\rho(\bm{X})\rho(\bm{X}_{\sigma\bm{\alpha}}^{-})\bm{v}(\bm{X})\cdot\bm{\alpha}d\Omega(\bm{\alpha})d\bm{X}\nonumber \\
 & -\frac{\sigma^{2}}{2m}\int g(\bm{X}_{\sigma\bm{\alpha}}^{+},\bm{X})\rho(\bm{X}_{\sigma\bm{\alpha}}^{+})\rho(\bm{X})\bm{v}(\bm{X})\cdot\bm{\alpha}d\Omega(\bm{\alpha})d\bm{X}\displaybreak[0]\nonumber \\
= & \frac{\sigma^{2}}{m}\int g(\bm{X},\bm{X}_{\sigma\bm{\alpha}}^{-})\rho(\bm{X})\rho(\bm{X}_{\sigma\bm{\alpha}}^{-})\bm{v}(\bm{X})\cdot\bm{\alpha}d\Omega(\bm{\alpha})d\bm{X}\nonumber \displaybreak[0] \\
=&- \frac{\sigma^{2}}{m}\int g(\bm{X},\bm{X}_{\sigma\bm{\alpha}}^{+})\rho(\bm{X})\rho(\bm{X}_{\sigma\bm{\alpha}}^{+})\bm{v}(\bm{X})\cdot\bm{\alpha}d\Omega(\bm{\alpha})d\bm{X},\label{eq:5.4}
\end{align}
\noindent
where $V_{\alpha}^{\prime}\equiv(\bm{\xi}_{*}^{\prime}-\bm{\xi}^{\prime})\cdot\bm{\alpha}=-V_{\alpha}$ has been used
at the beginning of the above transformation.

\subsection{Conservation laws and the collisional momentum and energy transports\label{app:ConservationLaws}}

Consider first the integration of (\ref{eq:2.1})
with respect to $\bm{\xi}$. 
Since $\langle J(f) \rangle=0$ by (\ref{eq:mom_J}) with ${\varphi}=1$, we have the standard form of the continuity equation:
\begin{subequations}
\begin{equation}
\frac{\partial\rho}{\partial t}+\frac{\partial\rho v_i}{\partial X_i}=0.\label{eq:continuity}
\end{equation}
\noindent
Next consider the integration of (\ref{eq:2.1}) multiplied by $\psi=\xi_j$ and $\psi=\bm{\xi}^2/2$ with respect to $\bm{\xi}$.
Then we have
\begin{align}
&\frac{\partial\rho v_j}{\partial t}
+\frac{\partial}{\partial X_i}(\rho v_i v_j+p_{ij}^{(k)})=\langle \xi_j J(f)\rangle,
\label{eq:momentum}
\\
&\frac{\partial}{\partial t}[\rho(e^{(k)}+\frac12 \bm{v}^2)] \notag\\
&+\frac{\partial}{\partial X_i}[\rho v_i(e^{(k)}+\frac12 \bm{v}^2)+p_{ij}^{(k)}v_j+q^{(k)}_i]=\frac12 \langle \bm{\xi}^2 J(f)\rangle,
\label{eq:energy}
\end{align}
\end{subequations}
\noindent
where the definitions \eqref{eq:kin_stress_def}, \eqref{eq:kin_heatfl_def}, and \eqref{eq:kin_inteng_def} have been used.

The difference from the Boltzmann equation occurs
in that there are contributions from collision term on the right-hand side.
Indeed, \eqref{eq:mom_J} with $\varphi=\psi$ leads to
\begin{subequations}\label{eq:4.3}
\begin{align}
  \langle\psi J(f) \rangle  
= &-\frac{\sigma^{2}}{m} \int(\psi-\psi^{\prime})g(\bm{X}_{\sigma\bm{\alpha}}^{-},\bm{X})f_*(\bm{X}_{\sigma\bm{\alpha}}^{-})f(\bm{X}) \notag\\ &\qquad\quad\times V_{\alpha}\theta(V_{\alpha})d\Omega(\bm{\alpha})d\bm{\xi}d\bm{\xi}_{*}.
\end{align}
But this expression can be further transformed
by using the operations (I) and (II) first 
and then using the relation $\psi$+$\psi_{*}$=$\psi^{\prime}$+$\psi_{*}^{\prime}$ as
\begin{align}
    \langle\psi J(f) \rangle
= &\frac{\sigma^{2}}{m} \int(\psi_{*}^{\prime}-\psi_{*})g(\bm{X}_{\sigma\bm{\alpha}}^{+},\bm{X})f(\bm{X}_{\sigma\bm{\alpha}}^{+})f_{*}(\bm{X}) \notag\\ &\qquad \times V_{\alpha}\theta(V_{\alpha})d\Omega(\bm{\alpha})d\bm{\xi}d\bm{\xi}_{*} \notag\\
= &\frac{\sigma^{2}}{m} \int(\psi-\psi^{\prime})g(\bm{X}_{\sigma\bm{\alpha}}^{+},\bm{X})f(\bm{X}_{\sigma\bm{\alpha}}^{+})f_{*}(\bm{X}) \notag\\ &\qquad\times V_{\alpha}\theta(V_{\alpha})d\Omega(\bm{\alpha})d\bm{\xi}d\bm{\xi}_{*}.
\label{eq:4.3a}
\end{align}
\noindent
Note that
\begin{equation}
\psi^{\prime}-\psi=\begin{cases}
V_{\alpha}\alpha_{i}, & {\displaystyle (\psi=\xi_{i}),}\\
{\displaystyle \frac{1}{2}V_{\alpha}(\bm{\xi}+\bm{\xi}_{*})\cdot\bm{\alpha},} & ({\displaystyle \psi=\frac{1}{2}\bm{\xi}^{2})},
\end{cases}\label{psi_p}
\end{equation}
\end{subequations}
\noindent
holds, thanks to the relation \eqref{eq:2.5}.
Hence, once integrated over the domain $D$,  
the result is expressed after the shift operation as
\begin{align}
    &\int_D \langle\psi J(f) \rangle d\bm{X} \notag\\
= &-\frac{\sigma^{2}}{m} \int(\psi-\psi^{\prime})g(\bm{X}_{\sigma\bm{\alpha}}^{-},\bm{X})f_*(\bm{X}_{\sigma\bm{\alpha}}^{-})f(\bm{X}) \notag\\&\qquad \times V_{\alpha}\theta(V_{\alpha})d\Omega(\bm{\alpha})d\bm{\xi}d\bm{\xi}_{*} d\bm{X} \notag \\
= &-\frac{\sigma^{2}}{m} \int(\psi-\psi^{\prime})g(\bm{X}_{\sigma\bm{\alpha}}^{+},\bm{X})f(\bm{X}_{\sigma\bm{\alpha}}^{+})f_*(\bm{X}) \notag\\&\qquad \times V_{\alpha}\theta(V_{\alpha})d\Omega(\bm{\alpha})d\bm{\xi}d\bm{\xi}_{*} d\bm{X}\notag \\
=&-\int_D \langle\psi J(f) \rangle d\bm{X},
\end{align}
\noindent
where \eqref{eq:4.3a} has been used at the last equality.
This implies that
\begin{equation}
\int_D \langle \psi J(f) \rangle d\bm{X}=0,
\quad (\psi=\xi_j, \frac{\bm{\xi}^2}{2}),
\label{eq:totalmom_eng_transp}
\end{equation}
\noindent
and accordingly the integration in space over the domain of \eqref{eq:continuity}, \eqref{eq:momentum}, and \eqref{eq:energy}
leads to
\begin{subequations}\label{eq:totalcons}
\begin{align}
& \frac{d}{dt}\int_D \rho d\bm{X}=0, \label{eq:totalmass} \\
& \frac{d}{dt}\int_D \rho v_jd\bm{X}
=\int_{\partial D}p_{ij}^{(k)}n_idS,\label{eq:totalmom} \\
& \frac{d}{dt}\int_D [\rho (e^{(k)}+\frac12\bm{v}^2)] d\bm{X}
=\int_{\partial D}(p_{ij}^{(k)}v_j+q_i^{(k)})n_idS,\label{eq:totaleng}
\end{align}\end{subequations}
\noindent
for three typical cases discussed in Sec.~\ref{sec:mono}.

In the case (i), the surface integrals in \eqref{eq:totalcons} vanish because of the periodic condition.
Accordingly, in addition to the total mass $\int_D \rho d\bm{X}$, the total momentum $\int_D \rho v_j d\bm{X}$ and total energy $\int_D \rho [e+(1/2)\bm{v}^2] d\bm{X}$  are constant in time.

In the case (ii), because of the specular reflection condition,
the surface integral in \eqref{eq:totaleng} vanishes,
although that in \eqref{eq:totalmom} does not vanish in general.
However, the total mass $\int_D \rho d\bm{X}$ and the total energy $\int_D \rho [e+(1/2)\bm{v}^2] d\bm{X}$ are constant in time. 
The gas approaches to a resting state or a rigid rotational motion in accordance with the boundary shape (see Sec.~\ref{sec:Maxwellian}). 
In order to see it a little more closely, consider the conservation law of the angular momentum, which is obtained by multiplying \eqref{eq:momentum} 
with $\epsilon_{ijk}X_k$, where $\epsilon_{ijk}$ is the Eddington epsilon. 
After integrating over the domain $D$, we have eventually
\begin{equation}
 \frac{d}{dt}\Omega_i
=\int_{\partial D} \epsilon_{ijk}p_{jl}^{(k)}n_lX_k dS,\label{eq:Omega}
\end{equation}
\noindent
where $\Omega_i=\int_D \epsilon_{ijk} \rho v_j X_k d\bm{X}$ is the total angular momentum.
In deriving \eqref{eq:Omega}, 
$\int_D\epsilon_{ijk} X_k\langle \xi_j J(f)\rangle d\bm{X}$ has vanished, 
since $\epsilon_{ijk} X_k \xi_j$ is a summational invariant 
[see \cite{TT24} and \eqref{eq:mom_Jsum}]. 
Because of the specular reflection, $p_{jl}^{(k)}n_l$ has no components tangential to the boundary. Thus the integrand on the right-hand side
vanishes if the position vector $\bm{X}$ of a point on the surface
is perpendicular to the tangential direction of the boundary around
the axis of rotation.
As the result, if the shape of the boundary is symmetric
around a certain axis, the total angular momentum $\Omega$ around the same axis does not change in time. If not, at the stationary equilibrium state discussed in Sec.~\ref{sec:Maxwellian},
the rigid rotation conflicts with the impermeability of the boundary and the flow ought to vanish.

In the case (iii), the surface integrals in \eqref{eq:totalcons} do not vanish in general,
since there can be the friction and energy exchanges on the impermeable surface of a heat bath. Hence only the total mass is constant in time.

In the meantime, there is another way of transformation of
$\langle \psi J(f)\rangle$, which leads to the concept of the collisional contributions to the stress tensor and the heat-flow vector.
To see it, combine \eqref{eq:4.3a} and (\ref{eq:mom_J})
for ${\varphi}=\psi$. Then, we have 
\begin{align}
 &\langle\psi J(f) \rangle \notag\\
= & \frac{1}{2}\frac{\sigma^{2}}{m}\int(\psi^{\prime}-\psi)\{g(\bm{X}_{\sigma\bm{\alpha}}^{-},\bm{X})f_{*}(\bm{X}_{\sigma\bm{\alpha}}^{-})f(\bm{X})\nonumber \\
 & -g(\bm{X}_{\sigma\bm{\alpha}}^{+},\bm{X})f(\bm{X}_{\sigma\bm{\alpha}}^{+})f_{*}(\bm{X})\}V_{\alpha}\theta(V_{\alpha})d\Omega(\bm{\alpha})d\bm{\xi}_{*}d\bm{\xi}.\label{eq:4.3b}
\end{align}
\noindent
This form allows further local transformation, thanks to \cite{CL88,MGB18}{,}
\begin{align}
 & g(\bm{X}_{\sigma\bm{\alpha}}^{-},\bm{X})f_{*}(\bm{X}_{\sigma\bm{\alpha}}^{-})f(\bm{X})-g(\bm{X}_{\sigma\bm{\alpha}}^{+},\bm{X})f(\bm{X}_{\sigma\bm{\alpha}}^{+})f_{*}(\bm{X})\displaybreak[0]\notag\\
= & -\int_{0}^{\sigma}\frac{\partial}{\partial\lambda}[g(\bm{X}_{\lambda\bm{\alpha}}^{+},\bm{X}_{(\lambda-\sigma)\bm{\alpha}}^{+})f_{*}(\bm{X}_{(\lambda-\sigma)\bm{\alpha}}^{+})f(\bm{X}_{\lambda\bm{\alpha}}^{+})]d\lambda\displaybreak[0]\notag\\
= & -\frac{\partial}{\partial X_i}\int_{0}^{\sigma}\alpha_i
 g(\bm{X}_{\lambda\bm{\alpha}}^{+},\bm{X}_{(\lambda-\sigma)\bm{\alpha}}^{+}) \notag\\ &\qquad\qquad\times f_{*}(\bm{X}_{(\lambda-\sigma)\bm{\alpha}}^{+})f(\bm{X}_{\lambda\bm{\alpha}}^{+})d\lambda.\label{eq:15}
\end{align}
\noindent
Then, eventually the right-hand side of \eqref{eq:momentum} and \eqref{eq:energy}
can be rewritten as \cite{CL88,MGB18}
\begin{subequations}
\begin{align}
 \langle\xi_j J(f) \rangle
= &-\frac{\partial}{\partial X_i}p_{ij}^{(c)}, \notag \\
  \frac{1}{2}\langle \bm{\xi}^{2}J(f)\rangle
=&-\frac{\partial}{\partial X_{i}}(p_{ij}^{(c)}v_{j}+q_{i}^{(c)}),\label{eq:4.8}
\end{align}
\noindent
where
\begin{align}
p_{ij}^{(c)}= & \frac{\sigma^{2}}{2m}\int {\int_{0}^{\sigma}}\alpha_{i}\alpha_{j}V_{\alpha}^{2}\theta(V_{\alpha}) g(\bm{X}_{\lambda\bm{\alpha}}^{+},\bm{X}_{(\lambda-\sigma)\bm{\alpha}}^{+}) \notag\\&\quad\times f_{*}(\bm{X}_{(\lambda-\sigma)\bm{\alpha}}^{+})f(\bm{X}_{\lambda\bm{\alpha}}^{+}){d\lambda} d\Omega(\bm{\alpha})d\bm{\xi}_{*}d\bm{\xi}, \label{eq:col_stress}\\
q_{i}^{(c)}= & \frac{\sigma^{2}}{4m}\int{\int_{0}^{\sigma}}\alpha_{i}[(\bm{c}+\bm{c}_{*})\cdot\bm{\alpha}]V_{\alpha}^{2}\theta(V_{\alpha}) \notag\\&\quad\times g(\bm{X}_{\lambda\bm{\alpha}}^{+},\bm{X}_{(\lambda-\sigma)\bm{\alpha}}^{+})f_{*}(\bm{X}_{(\lambda-\sigma)\bm{\alpha}}^{+}) \notag\\
&\qquad\times f(\bm{X}_{\lambda\bm{\alpha}}^{+}){d\lambda} d\Omega(\bm{\alpha})d\bm{\xi}_{*}d\bm{\xi},\label{eq:4.6}
\end{align}\end{subequations}
\noindent
with $\bm{c}=\bm{\xi}-\bm{v}$ and $\bm{c}_{*}=\bm{\xi}_{*}-\bm{v}$;
see, e.g., \cite{CL88,F99}.
Hence, \eqref{eq:momentum} and \eqref{eq:energy} are rewritten as
\begin{subequations}
\begin{align}
&\frac{\partial\rho v_j}{\partial t}
+\frac{\partial}{\partial X_i}(\rho v_i v_j+p_{ij})
=0,
\label{eq:mom_div}
\\
&\frac{\partial}{\partial t}[\rho(e+\frac12 \bm{v}^2)]
+\frac{\partial}{\partial X_i}[\rho v_i(e+\frac12 \bm{v}^2)
+p_{ij}v_j+q_i]=0,
\label{eq:eng_div}
\end{align}
\noindent
with
\begin{equation}
p_{ij}=p_{ij}^{(k)}+p_{ij}^{(c)},\quad
q_i=q_i^{(k)}+q_i^{(c)},\quad
e=e^{(k)},\label{eq:kin+col}
\end{equation}
\end{subequations}
\noindent
and
the usual form of the conservation laws of mass, momentum, and energy
has been recovered, together with \eqref{eq:continuity}.
The superscript $(c)$ denotes the collisional contribution to individual quantities.

\section{The kinetic boundary condition for the heat bath
and the Darrozes--Guiraud inequality \label{sec:KBC}}

The kinetic boundary condition on the impermeable surface $\partial D$
of a heat bath can be written generically as
\begin{subequations}\label{KBC}
\begin{align}
    f(t,\bm{X},\bm{\xi})=\int_{\bm{\xi}_{*}\cdot\bm{n}<0}
    & K(\bm{\xi},\bm{\xi}_{*}|\bm{X})f(t,\bm{X},\bm{\xi}_{*})d\bm{\xi}_{*}, \notag\\
    & \quad(\bm{\xi}\cdot\bm{n}>0,\ \bm{X}\in\partial D),    
\label{eq:2.10}
\end{align}
\noindent
where $K(\bm{\xi},\bm{\xi}_{*}|\bm{X})$ is a scattering kernel assumed
to be time-independent.
Assuming that the boundary is at rest,
the following properties are conventionally required to hold: \cite{S07}
\begin{enumerate}
\item Non-negativeness:
\begin{equation}
K(\bm{\xi},\bm{\xi}_{*}|\bm{X})\ge0,\quad(\bm{\xi}\cdot\bm{n}>0,\ \bm{\xi}_{*}\cdot\bm{n}<0);
\end{equation}
\item Normalization:
\begin{equation}
\int_{\bm{\xi}\cdot\bm{n}>0}\Big|\frac{\bm{\xi}\cdot\bm{n}}{\bm{\xi}_{*}\cdot\bm{n}}\Big|K(\bm{\xi},\bm{\xi}_{*}|\bm{X})d\bm{\xi}=1,\quad(\bm{\xi}_{*}\cdot\bm{n}<0),\label{eq:3.12}
\end{equation}
\noindent
where the integrand in (\ref{eq:3.12}) is the so-called reflection
probability density. Equation~(\ref{eq:3.12}) implies that the boundary
$\partial D$ is impermeable;
\item Preservation of equilibrium: The resting Maxwellian $f_{w}$ characterized by the surface temperature $T_{w}$
and defined by \eqref{eq:fw} satisfies the boundary condition (\ref{eq:2.10}),
and the other Maxwellians do not satisfy (\ref{eq:2.10}).
\end{enumerate}
\end{subequations}
\noindent
The diffuse reflection, the Maxwell, and the Cercignani--Lampis
condition \cite{CL71,C88,S07} that are widely used for the Boltzmann
equation are specific examples of \eqref{KBC}. Note that the uniqueness in the third
property listed above excludes the adiabatic boundary such as the
specular reflection condition.
On the boundary with the property \eqref{KBC},
the following statement is known to hold.
\begin{lemma}
\label{lem:(Darrozes--Guiraud)}(Darrozes--Guiraud \cite{DG66,C88,S07})
If the velocity distribution function $f$ satisfies the boundary
condition \eqref{KBC}, then it holds that
\begin{equation}
\int_{\partial D}\langle(\bm{\xi}\cdot\bm{n})f\ln\frac{f}{f_{w}}\rangle dS\le0.
\end{equation}
\noindent
Here the equality holds if and only if $f=f_{w}$.
\end{lemma}

\section{Boundedness of $\mathcal{H}$ and $\mathcal{F}$\label{app:boundedness}}

In this Appendix, we will show that $\mathcal{H}$ and $\mathcal{F}$ 
are bounded.

First consider the part $\mathcal{H}^{(k)}$. The following discussion on this part is along the same line as the proof in \cite{CIP94} for the Boltzmann equation.
We will show that $\mathcal{H}^{(k)}$ is bounded from below. 
To this end, it is enough to consider the case $f<1$,
since $\mathcal{H}^{(k)}=\int_D \int f\ln f d\bm{\xi}d\bm{X}$.
Separate the range of integration into three parts:
$\mathcal{D}^+=\{(\bm{x},\bm{\xi})|f\ge 1\}$, 
$\mathcal{D}^-_1=\{(\bm{x},\bm{\xi})|f< 1, f>\exp(-\beta\bm{\xi}^2/2) \}$, 
$\mathcal{D}^-_2=\{(\bm{x},\bm{\xi})|f< 1, f\le\exp(-\beta\bm{\xi}^2/2) \}$.
Here $\beta$ is a constant that is strictly positive.
Then, the contribution from the range $\mathcal{D}^+$ is non-negative,
\begin{equation}
\mathcal{H}^{(k)}\ge (\int_{\mathcal{D}^-_1}+\int_{\mathcal{D}^-_2}) f\ln f d\bm{\xi}d\bm{X}.\label{eq:estH^k}
\end{equation}
\noindent
The function $f\ln f$ decreases monotonically from $0$ to $-e^{-1}$ as $f$ increases from $0$ to $e^{-1}$, while $f\ln f>-f$ for $f>e^{-1}$. 
Therefore, 
$f\ln f>-(\beta\bm{\xi}^2/2)f$ in the range $\mathcal{D}^-_1$,
while $f\ln f\ge -f-(\beta\bm{\xi}^2/2)\exp(-\beta\bm{\xi}^2/2)$ in the range $\mathcal{D}^-_2$.
Applying these estimates to \eqref{eq:estH^k}, 
\begin{align}
\mathcal{H}^{(k)}
\ge &-\int_{\mathcal{D}^-_1} \frac{\beta\bm{\xi}^2}{2}f d\bm{\xi}d\bm{X} \notag\\
     &-\int_{\mathcal{D}^-_2} [f+\frac{\beta\bm{\xi}^2}{2}\exp(-\frac{\beta\bm{\xi}^2}{2})] d\bm{\xi}d\bm{X} \notag\\
\ge &-\int_D \langle \frac{\beta\bm{\xi}^2}{2}f \rangle d\bm{X}
     -\int_D  \langle f+\frac{\beta\bm{\xi}^2}{2}\exp(-\frac{\beta\bm{\xi}^2}{2}) \rangle d\bm{X} \notag\\
= & -\beta \int_D \rho(e^{(k)}+\frac12\bm{v}^2)d\bm{X}
    -\int_D \rho d\bm{X}+\mathrm{const.},\label{eq:H^kbound}
\end{align}
\noindent
where the last constant depends on $\beta$.
When the domain is periodic or is surrounded by specular reflection boundary,
both the total mass $\int_D \rho d\bm{X}$ and energy $\int_D \rho(e^{(k)}+\frac12\bm{v}^2)d\bm{X}$ are constant in time.
Hence $\mathcal{H}^{(k)}$ is bounded from below.
When the domain is surrounded by the impermeable surface of a resting heat bath with a uniform constant temperature $T_w$, put $\beta=(RT_w)^{-1}$.
Then, \eqref{eq:H^kbound} implies that $RT_w(\mathcal{H}^{(k)}+\int_D\langle(\bm{\xi}^2/2)f\rangle d\bm{X})$, in place of $\mathcal{H}^{(k)}$, is bounded from below.

Next consider the part $\mathcal{H}^{(c)}$ for the van der Waals case \eqref{eq:SvdW}
and for the Carnahan--Starling case \eqref{eq:SCS}.
For the van der Waals case, $\int_0^x\mathcal{S}(y)dy=-\ln(2-x)+\ln2\ge0$, so that $\mathcal{H}^{(c)}\ge0$ and is bounded from below.
For Carnahan--Starling case, $\mathcal{H}^{(c)}$ is again bounded from below, since $\int_0^x\mathcal{S}(y)dy=16(12-x)/(8-x)^2-3\ge0$.

To summarize, $\mathcal{H}$ (or $\mathcal{F}$) is bounded from below.
Hence, if $\mathcal{H}$ (or $\mathcal{F}$) is bounded at the initial time, it is also bounded entirely in time and approaches a stationary value as $t\to\infty$. 

\section{Extension to the Enskog--Vlasov equation\label{app:EV}}

In the case of the Enskog--Vlasov equation, an external force term $F_{i}{\partial f}/{\partial\xi_{i}}$ is added on the left-hand side of \eqref{eq:2.1}, where
\begin{equation}
F_{i}=-\int_{D}\frac{\partial}{\partial X_{i}}\Phi(|\bm{Y}-\bm{X}|)\rho(\bm{Y})d\bm{Y},\label{Vlasov}
\end{equation}
\noindent
and $\Phi$ is the attractive isotropic force potential between molecules. 

Since the $(1+{\ln{f}})$-moment of the external force term vanishes as
\begin{equation}
\langle(1+{\ln f})F_{i}\frac{\partial f}{\partial\xi_{i}}\rangle=\langle F_{i}\frac{\partial}{\partial\xi_{i}}(f{\ln f})\rangle=0,
\end{equation}
\noindent
the external force term does not contribute to (\ref{eq:27}).
Hence, \eqref{eq:H^k}, \eqref{eq:H_inequality}, and eventually \eqref{eq:H_ineq_period} for the periodic domain and \eqref{eq:H_ineq_spec} for the domain surrounded by the specular reflection boundary remain unchanged.

Next consider the multiplication of \eqref{eq:2.1} with $(1+\ln({f}/f_{w}))$ in the case of the domain surrounded by the impermeable surface of a heat bath with a uniform constant temperature $T_w$:
\begin{align}
\langle(1+\ln\frac{f}{f_{w}})F_{i}\frac{\partial f}{\partial\xi_{i}}\rangle &=-\langle({\ln{f_{w}}})F_{i}\frac{\partial f}{\partial\xi_{i}}\rangle \notag\\
&=F_{i}\langle\frac{\bm{\xi}^{2}}{2RT_{w}}\frac{\partial f}{\partial\xi_{i}}\rangle \notag\\
&=-\frac{\rho v_{i}F_{i}}{RT_{w}}.
\label{eq:externalF}
\end{align}
\noindent
Since $F_i$ is given by \eqref{Vlasov}, 
\begin{align}
&-\int_{D}\frac{\rho v_{i}F_{i}}{RT_{w}}d\bm{X} \notag\\
=& \int_{D}\frac{\rho v_{i}}{RT_{w}}\frac{\partial}{\partial X_{i}}\int_{D}\Phi(|\bm{Y}-\bm{X}|)\rho(\bm{Y})d\bm{Y}d\bm{X}\displaybreak[0]\notag\\
=&-\int_{\partial D}\frac{\rho v_{i}}{RT_{w}}n_{i}\int_{D}\Phi(|\bm{Y}-\bm{X}|)\rho(\bm{Y})d\bm{Y}dS\notag\\
&-\int_{D}\frac{1}{RT_{w}}\frac{\partial(\rho v_{i})}{\partial X_{i}}\int_{D}\Phi(|\bm{Y}-\bm{X}|)\rho(\bm{Y})d\bm{Y}d\bm{X}\displaybreak[0]\notag\\
=& \int_{D}\frac{1}{RT_{w}}\frac{\partial\rho(\bm{X})}{\partial t}\int_{D}\Phi(|\bm{Y}-\bm{X}|)\rho(\bm{Y})d\bm{Y}d\bm{X}\displaybreak[0]\notag\\
=& \frac{1}{2}\frac{d}{dt}\int_{D\times D}\frac{\Phi(|\bm{Y}-\bm{X}|)}{RT_{w}}\rho(\bm{X})\rho(\bm{Y})d\bm{X}d\bm{Y},
\end{align}
\noindent
where $v_in_i=0$ on $\partial D$ and the continuity equation \eqref{eq:continuity} have been used.
This result implies that, 
in place of $\mathcal{H}$ [or $\mathcal{F}$ defined by \eqref{eq:freeEng}],
\begin{subequations}
\begin{equation}
\widetilde{\mathcal{H}}\equiv \mathcal{H}+\frac{1}{2}\int_{D\times D}\frac{\Phi(|\bm{Y}-\bm{X}|)}{RT_{w}}\rho(\bm{X})\rho(\bm{Y})d\bm{X}d\bm{Y},
\end{equation}
\noindent
or
\begin{equation}
\widetilde{\mathcal{F}}\equiv \mathcal{F}+\frac{1}{2}\int_{D\times D}\Phi(|\bm{Y}-\bm{X}|)\rho(\bm{X})\rho(\bm{Y})d\bm{X}d\bm{Y},
\end{equation}
\end{subequations}
\noindent
decreases monotonically in time in the case of the Enskog--Vlasov {version of the EESM}.

The contribution of the Vlasov term to the momentum conservation 
is written as
\begin{equation}
\langle \xi_j F_{i}\frac{\partial f}{\partial\xi_{i}}\rangle
=-\rho F_{j}
=\rho(\bm{X}) \int_{D}\frac{\partial}{\partial X_{j}}\Phi(|\bm{Y}-\bm{X}|)\rho(\bm{Y})d\bm{Y},
\end{equation}
\noindent
for $\bm{X}\in D$.
Since $\bm{Y}$ can be expressed in two ways $\bm{Y}=\bm{X}\pm\bm{r}$,
it follows that
\begin{widetext}
\begin{align}
 &\rho(\bm{X}) \int_{D}\frac{\partial}{\partial X_{j}}\Phi(|\bm{Y}-\bm{X}|)\rho(\bm{Y})d\bm{Y} \notag \\
=&\frac12 \int \frac{r_j}{|\bm{r}|}\Phi^\prime(|\bm{r}|)
[\rho(\bm{X})\rho(\bm{X}_{\bm{r}}^{-})\chi_D(\bm{X}_{\bm{r}}^{-})\chi_D(\bm{X})
-\rho(\bm{X})\rho(\bm{X}_{\bm{r}}^{+})\chi_D(\bm{X}_{\bm{r}}^{+})\chi_D(\bm{X})]
 d\bm{r} \notag \displaybreak[0]\\
=&-\frac12 \int \frac{r_j}{|\bm{r}|}\Phi^\prime(|\bm{r}|)
\int_0^1 \frac{\partial}{\partial \lambda}[\rho(\bm{X}^+_{(\lambda-1)\bm{r}})\chi_D(\bm{X}^+_{(\lambda-1)\bm{r}})
\rho(\bm{X}^+_{\lambda\bm{r}})\chi_D(\bm{X}^+_{\lambda\bm{r}})]
d\lambda d\bm{r} \notag \displaybreak[0] \\
=&-\frac12 \int \frac{r_j}{|\bm{r}|}\Phi^\prime(|\bm{r}|)
\int_0^1 r_i\frac{\partial}{\partial X_i}[\rho(\bm{X}^+_{(\lambda-1)\bm{r}})\chi_D(\bm{X}^+_{(\lambda-1)\bm{r}})
\rho(\bm{X}^+_{\lambda\bm{r}})\chi_D(\bm{X}^+_{\lambda\bm{r}})]
d\lambda d\bm{r} \notag \displaybreak[0] \\
=&-\frac12 \frac{\partial}{\partial X_i}
\int \frac{r_ir_j}{|\bm{r}|}\Phi^\prime(|\bm{r}|)
\int_0^1 \rho(\bm{X}^+_{(\lambda-1)\bm{r}})\chi_D(\bm{X}^+_{(\lambda-1)\bm{r}})
\rho(\bm{X}^+_{\lambda\bm{r}})\chi_D(\bm{X}^+_{\lambda\bm{r}})
d\lambda d\bm{r}, 
\end{align}
\end{widetext}
\noindent
where $\Phi^\prime(x)=d\Phi(x)/dx$, which is non-negative because $\Phi$ is the attractive potential.

Similarly,
the contribution to the energy conservation is written as
\begin{align}
&\langle\frac{1}{2}\bm{\xi}^{2}F_{i}\frac{\partial f}{\partial\xi_{i}}\rangle \notag\\
=& -\rho v_{i}F_{i} \notag \\
=&\rho(\bm{X}) v_i(\bm{X})
  \int_{D}\frac{\partial}{\partial X_{i}}\Phi(|\bm{Y}-\bm{X}|)\rho(\bm{Y})d\bm{Y} \notag\displaybreak[0] \\
=&\frac{\partial}{\partial X_{i}}
[\rho(\bm{X}) v_i(\bm{X})
  \int_{D}\Phi(|\bm{Y}-\bm{X}|)\rho(\bm{Y})d\bm{Y}] \notag\\
 &-\frac{\partial\rho v_i}{\partial X_{i}}
  \int_{D}\Phi(|\bm{Y}-\bm{X}|)\rho(\bm{Y})d\bm{Y}
  \notag \displaybreak[0]\\
=&\frac{\partial}{\partial X_{i}}
[\rho(\bm{X}) v_i(\bm{X})
  \int_{D}\Phi(|\bm{Y}-\bm{X}|)\rho(\bm{Y})d\bm{Y}] \notag\\
 &+\frac{\partial\rho(\bm{X})}{\partial t}
  \int_{D}\Phi(|\bm{Y}-\bm{X}|)\rho(\bm{Y})d\bm{Y}
  \notag \displaybreak[0]\\
=&\frac{\partial}{\partial X_{i}}
[\rho(\bm{X}) v_i(\bm{X})
  \int_{D}\Phi(|\bm{Y}-\bm{X}|)\rho(\bm{Y})d\bm{Y}] \notag\\
& +\frac12\int_{D}\Phi(|\bm{Y}-\bm{X}|)[\frac{\partial\rho(\bm{X})}{\partial t}\rho(\bm{Y})-\frac{\partial\rho(\bm{Y})}{\partial t}\rho(\bm{X})]d\bm{Y}
  \notag \\
& +\frac12\int_{D}\Phi(|\bm{Y}-\bm{X}|)[\frac{\partial\rho(\bm{X})}{\partial t}\rho(\bm{Y})+\frac{\partial\rho(\bm{Y})}{\partial t}\rho(\bm{X})]d\bm{Y}
  \notag\displaybreak[0] \\
=&\frac{\partial}{\partial X_{i}}
[\rho(\bm{X}) v_i(\bm{X})
  \int_{D}\Phi(|\bm{Y}-\bm{X}|)\rho(\bm{Y})d\bm{Y}] \notag\\
& +\frac12\int_{\mathbb{R}^3} \Phi(|\bm{r}|)[\frac{\partial\rho(\bm{X})}{\partial t}\rho(\bm{X}^+_{\bm{r}})\chi_D(\bm{X})\chi_D(\bm{X}^+_{\bm{r}}) \notag\\&\qquad-\frac{\partial\rho(\bm{X}^-_{\bm{r}})}{\partial t}\rho(\bm{X})\chi_D(\bm{X})\chi_D(\bm{X}^-_{\bm{r}})]d\bm{r}
  \notag \\
 & +\frac12\frac{\partial}{\partial t}
  \int_{D}\Phi(|\bm{Y}-\bm{X}|)\rho(\bm{X})\rho(\bm{Y})d\bm{Y}
  \notag \displaybreak[0]\\
=&\frac{\partial}{\partial X_{i}}
[\rho(\bm{X}) v_i(\bm{X})
  \int_{D}\Phi(|\bm{Y}-\bm{X}|)\rho(\bm{Y})d\bm{Y}] \notag\displaybreak[0]\\
& +\frac12\int_{\mathbb{R}^3} \int_0^1\Phi(|\bm{r}|)\frac{\partial}{\partial\lambda}
[\frac{\partial\rho(\bm{X}^+_{(\lambda-1)\bm{r}})}{\partial t}\rho(\bm{X}^+_{\lambda\bm{r}}) \notag\\&\quad\times \chi_D(\bm{X}^+_{(\lambda-1)\bm{r}})\chi_D(\bm{X}^+_{\lambda\bm{r}})]d\lambda d\bm{r}
  \notag \displaybreak[0]\\
 & +\frac12\frac{\partial}{\partial t}
  \int_{D}\Phi(|\bm{Y}-\bm{X}|)\rho(\bm{X})\rho(\bm{Y})d\bm{Y}
  \notag\displaybreak[0] \\
=&\frac{\partial}{\partial X_{i}}
[\rho(\bm{X}) v_i(\bm{X})
  \int_{D}\Phi(|\bm{Y}-\bm{X}|)\rho(\bm{Y})d\bm{Y} \notag\\
& +\frac12\int_{\mathbb{R}^3} \int_0^1 r_i \Phi(|\bm{r}|)
\frac{\partial\rho(\bm{X}^+_{(\lambda-1)\bm{r}})}{\partial t}\rho(\bm{X}^+_{\lambda\bm{r}}) \notag\\&\qquad\times\chi_D(\bm{X}^+_{(\lambda-1)\bm{r}})\chi_D(\bm{X}^+_{\lambda\bm{r}}) d\lambda d\bm{r} ]
  \notag \\
 & +\frac12\frac{\partial}{\partial t}
  \int_{D}\Phi(|\bm{Y}-\bm{X}|)\rho(\bm{X})\rho(\bm{Y})d\bm{Y}.
\end{align}
\noindent
These results lead to the concept of additional contributions
to the stress tensor, the heat-flow vector, and the internal energy  from the Vlasov term. If respectively denoted by $p_{ij}^{(v)}$,  $q_i^{(v)}$, and $e^{(v)}$, they are expressed as follows:
\begin{subequations}
\begin{align}
 p_{ij}^{(v)}=&-\frac12
\int_{\mathbb{R}^3} \frac{r_ir_j}{|\bm{r}|}\Phi^\prime(|\bm{r}|)
\int_0^1 \rho(\bm{X}^+_{\lambda\bm{r}})\chi_D(\bm{X}^+_{\lambda\bm{r}}) \notag\\&\qquad\times \rho(\bm{X}^+_{(\lambda-1)\bm{r}})\chi_D(\bm{X}^+_{(\lambda-1)\bm{r}})
d\lambda d\bm{r}, \label{eq:pijV}\displaybreak[0]\\
 q_{i}^{(v)}=&\frac12
\int_{\mathbb{R}^3} r_i\Phi(|\bm{r}|)
\int_0^1 \frac{\partial\rho(\bm{X}^+_{(\lambda-1)\bm{r}})}{\partial t}\chi_D(\bm{X}^+_{(\lambda-1)\bm{r}}) \notag\\&\qquad\times
\rho(\bm{X}^+_{\lambda\bm{r}})\chi_D(\bm{X}^+_{\lambda\bm{r}})
d\lambda d\bm{r} \notag\\
&+\frac12 v_j\int_{\mathbb{R}^3} \int \frac{r_ir_j}{|\bm{r}|}\Phi^\prime(|\bm{r}|)
\int_0^1 \rho(\bm{X}^+_{\lambda\bm{r}})\chi_D(\bm{X}^+_{\lambda\bm{r}}) \notag\\&\qquad\times \rho(\bm{X}^+_{(\lambda-1)\bm{r}})\chi_D(\bm{X}^+_{(\lambda-1)\bm{r}})
d\lambda d\bm{r} \notag \\
&+\frac12 \rho v_i \int_D \Phi(|\bm{X}-\bm{Y}|)\rho(\bm{Y})d\bm{Y}, \displaybreak[0]\\
 e^{(v)}=&\frac{1}{2}\int_D \Phi(|\bm{Y}-\bm{X}|)\rho(\bm{Y})d\bm{Y}.
\end{align}
\end{subequations}
\noindent
Hence, 
by redefining the stress tensor, heat-vector, and {specific} internal energy as
\begin{align}
&p_{ij}=p_{ij}^{(k)}+p_{ij}^{(c)}+p_{ij}^{(v)},\quad
q_i=q_i^{(k)}+q_i^{(c)}+q_i^{(v)},\notag\\
&e=e^{(k)}+e^{(v)},\label{eq:kin+col+vl}
\end{align}
\noindent
\eqref{eq:mom_div} and \eqref{eq:eng_div} are recovered and form the usual system of the conservation equations, together with the continuity equation \eqref{eq:continuity} that remains unchanged.
\cite{fn:anglmom}

In the uniform equilibrium state in the bulk,
$p_{ij}^{(v)}$ and $e^{(v)}$ are reduced to
\begin{subequations}
\begin{align}
& p_{ij}^{(v)}=-\frac{2\pi}{3}\rho^2
\int_0^\infty {x}^3 \Phi^\prime(x) dx\delta_{ij},\displaybreak[0] \\
& e^{(v)}=2\pi\rho \int_0^\infty x^2\Phi(x) dx,
\end{align}
\end{subequations}
\noindent
and thus the attractive part $p^{(v)}$ defined by
\begin{equation}
p^{(v)}=-a\rho^2,\quad
a\equiv \frac{2\pi}{3} \int_0^\infty x^3 \Phi^\prime(x) dx (>0),
\end{equation}
\noindent
is added to the right-hand side of the {EoS},
e.g., \eqref{eq:EoSvdW} and \eqref{eq:EoSCS}.
In this way, the attractive part of the {EoS}, if exists,
is recovered by the Vlasov term.
Incidentally, as far as $x^3\Phi(x)=0$ for $x=0$ and $x\to\infty$,
$e^{(v)}$ can be rewritten as $e^{(v)}=-a\rho$
and this is consistent with the equilibrium-thermodynamic relation
$e=e_\mathrm{ideal}+\int (p-T\partial p/\partial T)/\rho^2 d\rho$,
where $e_\mathrm{ideal}=(3/2)RT$ is the internal energy of ideal monatomic gases.

Finally, if $\Phi$ is bounded from below, it holds that
\begin{equation}
\int_{D\times D} \Phi(|\bm{Y}-\bm{X}|)\rho(\bm{X})\rho(\bm{Y}) d\bm{X} d\bm{Y}\ge C(\int_D \rho d\bm{X})^2,
\end{equation}
\noindent
where $C$ is a certain constant. Since $\int_D \rho d\bm{X}$ is the total mass, it is conserved in time. Thus, the boundedness of $\mathcal{H}$ and $\mathcal{F}$ shown in Appendix~\ref{app:boundedness} is extended to that
of  $\widetilde{\mathcal{H}}$ and $\widetilde{\mathcal{F}}$ in the case of the Enskog--Vlasov equation.

\acknowledgments
The research of the first author is supported in part by the JSPS Grant-in-Aid for Scientific Research({C}) (No.~22K03923). The second author is a JSPS Research Fellowship DC1 and his research is supported by the JSPS Grant-in-Aid for JSPS Fellows (No.~24KJ1450).


\begin{thebibliography}{99}
%
\bibitem{E72}
\newblock D. Enskog, 
\newblock Kinetic theory of heat conduction, viscosity,
and self-diffusion in compressed gases and liquids, 
\newblock in \emph{Kinetic Theory}, Vol. 3, S. G. Brush ed., Pergamon Press, Oxford, Part 2, 1972, pp.226--259.
%
{\bibitem{MS96}
\newblock J. M. Montanero and A. Santos,
\newblock Monte Carlo simulation method for the Enskog equation,
\newblock \emph{Phys. Rev. E} \textbf{54}, 438--444 (1996).}
%
\bibitem{F97} 
\newblock A. Frezzotti, 
\newblock \href{https://doi.org/10.1063/1.869247}{A particle scheme for the numerical solution of the Enskog equation}, 
\newblock \emph{Phys. Fluids} \textbf{9}, 1329--1335 (1997). 
%
{\bibitem{MS97a}
\newblock J. M. Montanero and A. Santos,
\newblock Simulation of the Enskog equation \emph{\`{a} la} Bird,
\newblock \emph{Phys. Fluids} \textbf{9}, 2057--2060 (1997).}
%
\bibitem{WLRZ16} 
\newblock L. Wu, H. Liu, J. M. Reese, and Y. Zhang,
\newblock \href{https://doi.org/10.1017/jfm.2016.173}{Non-equilibrium dynamics of dense gas under tight confinement}, 
\newblock \emph{J. Fluid Mech.} \textbf{794}, 252--266 (2016).
%
{\bibitem{F97b} 
\newblock A. Frezzotti, 
\newblock Molecular dynamics and Enskog theory calculation 
of one dimensional problems in the dynamics of dense gases, 
\newblock \emph{Physica A} \textbf{240}, 202--211 (1997).}
%
\bibitem{F99}
\newblock A. Frezzotti, 
\newblock \href{https://doi.org/10.1016/S0997-7546(99)80008-9}{Monte Carlo simulation of the heat flow in a dense hard sphere gas}, \newblock \emph{Eur. J. Mech. B/Fluids} \textbf{18}, 103--119 (1999).
%
\bibitem{FGLS18}
\newblock A. Frezzotti, L. Gibelli, D. A. Lockerby, and J. E. Sprittles,
\newblock \href{https://link.aps.org/doi/10.1103/PhysRevFluids.3.054001}{Mean-field kinetic theory approach to evaporation of a binary liquid into vacuum},
\newblock \emph{Phys. Rev. Fluids} \textbf{3}, 054001 (2018). 
%
\bibitem{KKW14}
\newblock M. Kon, K. Kobayashi, and M. Watanabe, 
\newblock \href{https://doi.org/10.1063/1.4890523}{Method of determining kinetic boundary conditions in net evaporation/condensation}, 
\newblock \emph{Phys. Fluids} \textbf{26}, 072003 (2014). 
%
\bibitem{HTT22}
\newblock M. Hattori, S. Tanaka and S. Takata,
\newblock  \href{https://doi.org/10.1063/5.0091390}{Heat transfer in a dense gas between two parallel plates}, 
\newblock \emph{AIP Advances} \textbf{12}, 055323 (2022).
%
\bibitem{VE73} 
\newblock H. van Beijeren and M. H. Ernst, 
\newblock \href{https://doi.org/10.1016/0031-8914(73)90372-8}{The modified Enskog equation}, 
\newblock \emph{Physica} \textbf{68}, 437--456 (1973). 
%
\bibitem{R78}
\newblock P. Resibois, 
\newblock \href{https://doi.org/10.1007/BF01011771}{H-theorem for the (modified) nonlinear Enskog equation},
\newblock \emph{J. Stat. Phys.} \textbf{19}, 593--609 (1978).
%
\bibitem{DVK21}
\newblock J. R. Dorfman, H. van Beijeren, and T. R. Kirkpatrick,
\newblock \emph{Contemporary Kinetic Theory of Matter}, 
\newblock Cambridge University Press, Cambridge, 2021.
%
\bibitem{BLPT91}
\newblock N. Bellomo, M. Lachowicz, J. Polewczak and G. Toscani,
\newblock \emph{Mathematical Topics in Nonlinear Kinetic Theory II},
\newblock World Scientific, Singapore, 1991. 
%
\bibitem{MGB18}
\newblock P. Maynar, M. I. Garcia de Soria, and J. J. Brey,
\newblock \href{https://doi.org/10.1007/s10955-018-1971-7}{The Enskog equation for confined elastic hard spheres}, 
\newblock \emph{J. Stat. Phys.} \textbf{170}, 999--1018 (2018). 
%
\bibitem{BB18}
\newblock E. S. Benilov and M. S. Benilov, 
\newblock \href{https://doi.org/10.1103/PhysRevE.97.062115}{Energy conservation and H theorem for the Enskog--Vlasov equation}, 
\newblock \emph{Phys. Rev. E} \textbf{97}, 062115 (2018). 
%
\bibitem{BB19} 
\newblock E. S. Benilov and M. S. Benilov,
\newblock The Enskog--Vlasov equation: a kinetic model describing gas, liquid, and solid,
\newblock \emph{J. Stat. Mech.} \textbf{2019}, 103205 (2019).
%
\bibitem{T24}
\newblock S. Takata, 
\newblock On the thermal relaxation of a dense gas described by the modified Enskog equation in a closed system in contact with a heat bath,
\newblock \emph{Kinetic \& Related Models} \textbf{17}, 331--346 (2024).
%
{\bibitem{MS97}
\newblock J. M. Montanero and A. Santos,
\newblock Viscometric effects in a dense hard-sphere fluid,
\newblock \emph{Physica A} \textbf{240}, 229--238 (1997).
%
\bibitem{G71}
\newblock M. Grmela, 
\newblock Kinetic equation approach to phase transitions,
\newblock \emph{J. Stat. Phys.} \textbf{3}, 347--364 (1971).
%
\bibitem{S16}
\newblock R. Soto,
\newblock \emph{Kinetic Theory and Transport Phenomena}, 
\newblock Oxford University Press, New York, 2016.
%
\bibitem{CS69}
\newblock N. F. Carnahan and K. E. Starling,
\newblock Equation of state for non-attracting rigid spheres,
\newblock \emph{J. Chem. Phys.} \textbf{51}, 635--636 (1969).
%
\bibitem{H23}
\newblock M. Hattori,
\newblock Poiseuille and thermal transpiration flows of a dense gas between two parallel plates,
\newblock \emph{J. Fluid Mech.} \textbf{962}, A20 (2023).}
%
\bibitem{fn1}
To be precise, it is necessary to make the argument of the logarithmic function dimensionless, like $\ln (f/c_0)$ with a constant $c_0$ having the same dimension as $f$.
We, however, leave the argument dimensional 
to avoid additional calculations that do not affect the results.
%
{\bibitem{fn:periodic}
Here $D$ should be considered as a unit cell of periodic domain.
The indicator function should be interpreted as $\chi_D(\bm{X})\equiv 1$
and the position $\bm{Y}$ is allowed to be in the next unit cell
so that the distance $|\bm{X}-\bm{Y}|$ occurring in $\theta(\sigma-|\bm{X}-\bm{Y}|)$
is measured naturally. 
%
\bibitem{DG66}
\newblock J. S. Darrozes and J. P. Guiraud, 
\newblock \href{https://gallica.bnf.fr/ark:/12148/bpt6k6238594s/f400.image.r=Darrouzes?rk=21459;2}{G\'{e}n\'{e}ralisation formelle du th\'{e}or\`{e}me H en pr\'{e}sence de parois. Applications},
\newblock \emph{C. R. Acad. Sci. Paris A} \textbf{262}, 1368--1371 (1966).
%
\bibitem{C88}
\newblock C. Cercignani, 
\newblock \emph{The Boltzmann Equation and Its Applications}, Springer, New York, 1988.
%
\bibitem{S07}
\newblock Y. Sone, 
\newblock \emph{Molecular Gas Dynamics}, Birkh\"{a}uer, Boston, 2007.
\newblock  Supplement is available from \href{http://hdl.handle.net/2433/66098}{http://hdl.handle.net/2433/66098}.}
%
\bibitem{fn:freeEng}
The factor $RT_w$ is multiplied in \eqref{eq:freeEng} 
for the correspondence to the thermodynamic free energy.
The last constant in \eqref{eq:freeEng} comes from the total mass conservation in the domain $D$.
%
{\bibitem{GG80}
\newblock M. Grmela and L. S. Garcia-Colin,
\newblock Compatibility of the Enskog kinetic theory with thermodynamics. I
\newblock \emph{Phys. Rev. A} \textbf{22}, 1295--1304 (1980).}
%
\bibitem{TT24}
\newblock S. Takata and A. Takahashi,
\newblock Note on the summational invariant and corresponding local Maxwellian for the Enskog equation,
\newblock {\emph{Kinetic \& Related Models}} \textbf{17}, 739--754 (2024).
%
{\bibitem{DSB96}
\newblock J. W. Dufty, A. Santos, and J. J. Brey,
\newblock Practical kinetic model for hard sphere dynamics,
\newblock \emph{Phys. Rev. Letters} \textbf{77}, 1270--1273 (1996).
%
\bibitem{MTHT23}
\newblock T. Miyauchi, S. Takata, M. Hattori, and A. Takahashi,
\newblock Constructions of simple kinetic equations for a dense gas,
\newblock \emph{Springer Proceedings in Mathematics \& Statistics} \textbf{429}, 19--39 (2023).
%
\bibitem{SSGZ23}
\newblock B. Shan, W. Su, L. Gibelli, and Y. Zhang,
\newblock Molecular kinetic modelling of non-equilibrium transport of confined van der Waals fluids,
\newblock \emph{J. Fluid Mech.} \textbf{976}, A7 (2023).
%
\bibitem{MT24}
\newblock T. Miyauchi and S. Takata,
\newblock Single droplet or bubble and its stability: Kinetic theory and dynamical system approaches,
\newblock \emph{Phys. Rev. E} \textbf{110}, 025102 (2024).}
%
\bibitem{vdW}
\newblock J. D. van der Waals,
\newblock Over de Continu\"{i}teit van den Gas -- en Vloeistoftoestand,
Academisch Proefschrift, Leiden (1873) (in Dutch);
\newblock see also the English translation: 
\newblock R. Threlfall and J.F. Adair,
\newblock \emph{Physical Memoirs} \textbf{1}, 333--496 (1890).
%
{\bibitem{CL88}
\newblock C. Cercignani and M. Lampis, 
\newblock \href{https://doi.org/10.1007/BF01014218}{On the kinetic theory of a dense gas of rough spheres},
\newblock \emph{J. Stat. Phys.} \textbf{53}, 655--672 (1988).}
%
\bibitem{CL71}
\newblock C. Cercignani and M. Lampis, 
\newblock \href{https://doi.org/10.1080/00411457108231440}{Kinetic models for gas--surface interactions}, 
\newblock \emph{Trans. Theory Stat. Phys.} \textbf{1}, 101--114  (1971). 
%
{\bibitem{CIP94}
\newblock C. Cercignani, R. Illner, and M. Pulvirenti,
\newblock \href{https://doi.org/10.1007/978-1-4419-8524-8}{\emph{The Mathematical Theory of Dilute Gases}}, Springer, New York, 1994, Sec.~9.4.}
%
\bibitem{fn:anglmom}
The discussions on the total angular momentum given in Appendix~\ref{app:ConservationLaws} can be extended straightforwardly to the case of the Enskog--Vlasov equation. 
In order to see it, consider $I_i\equiv \int_D\epsilon_{ijk}\langle \xi_j\partial(F_l f)/\partial \xi_l \rangle X_k d\bm{X}$. Since $F_i=-\int_D \rho(\bm{Y})[\partial \Phi(|\bm{X}-\bm{Y}|)/\partial X_i]d\bm{Y}$, 
$I_i$ is transformed after some manipulations into
$I_i=(\epsilon_{ijk}/2)\int_D\int_D \rho(\bm{X})\rho(\bm{Y})(X_k-Y_k)(\partial \Phi/\partial X_j) d\bm{X}d\bm{Y}$. Since $\partial \Phi/\partial X_j=\Phi^\prime(|\bm{X}-\bm{Y}|)(X_j-Y_j)/|\bm{X}-\bm{Y}|$, the integral is symmetric with respect to the indices $j$ and $k$.
Hence $I_i=0$ because of the multiplication with $\epsilon_{ijk}$, which means no contribution to the total angular momentum from the Vlasov term.
%
\end{thebibliography}

\end{document}